\documentclass[prd,showkeys,floatfix,twocolumn,amsmath,amssymb,floatfix]{revtex4}
\usepackage{graphicx}
\usepackage{float}
\usepackage{epstopdf}
\usepackage{multirow}
\usepackage{subfigure}
\usepackage{color}
\usepackage[colorlinks,citecolor=blue,anchorcolor=red,menucolor=red,linkcolor=red,filecolor=red,runcolor=red,urlcolor=blue,frenchlinks=red]{hyperref}

\allowdisplaybreaks[3]

\begin{document}
\title{Identifying the $\Xi_b(6100)$ as the $P$-wave bottom baryon of $J^P=3/2^-$}
%

\author{Hui-Min Yang$^1$}
\email{huiminy@buaa.edu.cn}
\author{Hua-Xing Chen$^2$}
\email{hxchen@seu.edu.cn}
\author{Er-Liang Cui$^3$}
\email{erliang.cui@nwafu.edu.cn}
\author{Qiang-Mao$^4$}
\email{maoqiang@ahszu.edu.cn}

\affiliation{
$^1$School of Physics, Beihang University, Beijing 100191, China \\
$^2$School of Physics, Southeast University, Nanjing 210094, China\\
$^3$College of Science, Northwest A$\&$F University, Yangling 712100, China\\
$^4$Department of Electrical and Electronic Engineering, Suzhou University, Suzhou 234000, China
}

\begin{abstract}
We study the $\Xi_b(6100)$ using the methods of QCD sum rules and light-cone sum rules within the framework of heavy quark effective theory. Our results suggest that the $\Xi_b(6100)$ can be well interpreted as the $P$-wave bottom baryon of $J^P=3/2^-$, belonging to the $SU(3)$ flavor $\mathbf{\bar 3}_F$ representation. It has a partner state of $J^P=1/2^-$, labelled as $\Xi_b(1/2^-)$, whose mass and width are predicted to be $m_{\Xi_b(1/2^-)}=6.08^{+0.13}_{-0.11}$~GeV and $\Gamma_{\Xi_b(1/2^-)}=4^{+29}_{-~4}$~MeV, with the mass splitting $\Delta M=m_{\Xi_b(6100)}-m_{\Xi_b(1/2^-)}=9\pm3$~MeV. We propose to search for it in the $\Xi_c({1/2}^-)\to \Xi_b^{\prime}\pi$ decay channel. Our results also suggest that the $\Lambda_b(5912)$ and $\Lambda_b(5920)$ are their partner states with $J^P=1/2^-$ and $3/2^-$ respectively, and moreover, the $\Lambda_c(2595)$, $\Lambda_c(2625)$, $\Xi_c(2790)$, and $\Xi_c(2815)$ are their charmed partner states.
\end{abstract}
\pacs{14.20.Mr, 12.38.Lg, 12.39.Hg}
\keywords{bottom baryon, heavy quark effective theory, QCD sum rules, light-cone sum rules}
\maketitle
\pagenumbering{arabic}
%
%
%
\section{Introduction}\label{sec:intro}
%

A singly heavy baryon consists of a heavy quark and two light quarks, where the light quarks together with gluons circle around the nearly static heavy quark. In the infinite $m_Q$ limit, the heavy quark is decoupled from the dynamics of the light quarks, which greatly simplifies a triquark system into a diquark system. This system is similar to the deuterium atom~\cite{Korner:1994nh,Manohar:2000dt,Bianco:2003vb,Klempt:2009pi,Kalman:1983an}. It is well known that the electromagnetic interaction produces the light spectroscopies with the fine splitting. Similarly, the strong interaction also produces the hadron spectroscopies with the fine splitting, and the singly heavy baryon system provides an ideal platform to study this effect~\cite{Isgur:1978wd,Copley:1979wj,Karliner:2008sv,Chen:2016spr,Chen:2022asf}.

In the past fifty years many important experimental progresses have been made in the field of singly heavy baryons. Since the lowest-lying charmed baryon $\Lambda_c$ was reported by Fermilab in 1976~\cite{Knapp:1976qw}, a lot of singly heavy baryons have been observed in experiments~\cite{pdg}, {\it e.g.}:
\begin{itemize}

\item The $\Lambda_c(2595)$, $\Lambda_c(2625)$, $\Xi_c(2790)$ and $\Xi_c(2815)$ can be well interpreted as the $P$-wave charmed baryons completing two flavor $\mathbf{\bar 3}_F$ multiplets of $J^P=1/2^-$ and $3/2^-$~\cite{ARGUS:1993vtm,E687:1993bax,CLEO:1994oxm,CLEO:1999msf}. Their masses and widths (or upper limits at 90\% credibility level) were measured to be~\cite{pdg}:
\begin{eqnarray}
\Lambda_c(2595)^+ &:& M = 2592.25 \pm 0.28 {\rm~MeV} \, ,
\\ \nonumber      && \Gamma = 2.59 \pm 0.30\pm 0.47{\rm~MeV} \, ;
\\ \Lambda_c(2625)^+ &:& M = 2628.11 \pm 0.19 {\rm~MeV} \, ,
\\ \nonumber      && \Gamma < 0.97{\rm~MeV}~{\rm at}~90\%~{\rm CL} \, ;
\\ \Xi_c(2790)^0   &:& M = 2793.9 \pm 0.5{\rm~MeV} \, ,
\\ \nonumber      && \Gamma = 10.0  \pm 0.7 \pm 0.8{\rm~MeV} \, ;
\\ \Xi_c(2815)^0    &:& M = 2819.79 \pm 0.30 {\rm~MeV} \, ,
\\ \nonumber      && \Gamma = 2.54  \pm 0.18 \pm 0.17{\rm~MeV} \, .
\end{eqnarray}

\item In 2012 the LHCb collaboration discovered two narrow states, the $\Lambda_b(5912)$ and $\Lambda_b(5920)$, in the $\Lambda_b\pi\pi$ mass spectrum~\cite{LHCb:2012kxf}. Their masses and widths were measured to be~\cite{pdg}:
\begin{eqnarray}
\Lambda_b(5912)^0 &:& M = 5912.19 \pm 0.17{\rm~MeV} \, ,
\\ \nonumber      && \Gamma < 0.25{\rm~MeV}~{\rm at}~90\%~{\rm CL} \, ;
\\ \Lambda_b(5920)^0 &:& M = 5920.09 \pm 0.17{\rm~MeV} \, ,
\\ \nonumber      && \Gamma < 0.19{\rm~MeV}~{\rm at}~90\%~{\rm CL} \, .
\end{eqnarray}

\item Recently, the $\Xi_b(6100)$ was observed by the CMS collaboration in the $\Xi_b^-\pi^+\pi^-$ invariant mass spectrum~\cite{CMS:2021rvl}. Its mass and width were measured to be:
\begin{eqnarray}
\Xi_b(6100)^- &:& M = 6100.3 \pm 0.2 \pm 0.1 \pm 0.6{\rm~MeV} \, ,
\\ \nonumber      && \Gamma <1.9{\rm~MeV}~{\rm at}~95\%~{\rm CL} \, .
\end{eqnarray}

\end{itemize}

Experimental progresses on the singly heavy baryons have attracted a lot of theorists to study them. It is a challenging issue to fully understand their internal structures. Various theoretical methods and models have been applied in this field, including various quark models~\cite{Roberts:2007ni,Bijker:2020tns,Chen:2018vuc,Chen:2018orb,Wang:2018fjm,Xiao:2020gjo,Ebert:2011kk,Chen:2016iyi,Yang:2018lzg,Lu:2020ivo,Wang:2020gkn,Wang:2019uaj}, the chiral perturbation theory~\cite{Lu:2014ina,Cheng:2015naa}, the chiral unitary approach~\cite{Zeng:2020och}, various molecular interpretations~\cite{Huang:2018bed,Liang:2017ejq,Chen:2017xat,Liang:2014eba,An:2017lwg,Debastiani:2017ewu}, the Regge trajectory~\cite{Guo:2008he}, the $^3P_0$ model~\cite{Chen:2007xf,Ye:2017yvl}, the relativistic flux tube model~\cite{Chen:2014nyo}, QCD sum rules~\cite{Aliev:2018vye,Wang:2020pri,Agaev:2020fut,Azizi:2020tgh,Yu:2021zvl}, and the lattice QCD~\cite{Burch:2015pka,Padmanath:2013bla,Padmanath:2017lng}, etc. We refer to the reviews~\cite{Crede:2013kia,Cheng:2015iom,Cheng:2021qpd,Meng:2022ozq,Chen:2016spr,Chen:2022asf} for detailed discussions.

In our previous works~\cite{Yang:2021lce,Yang:2020zrh,Yang:2020zjl,Mao:2020jln,Cui:2019dzj,Chen:2020mpy,Chen:2017sci,Mao:2015gya,Chen:2015kpa,Yang:2019cvw} we have studied the mass spectra and decay properties of the $P$-wave charmed and bottom baryons using the methods of QCD sum rules~\cite{Shifman:1978bx,Reinders:1984sr} and light-cone sum rules~\cite{Balitsky:1989ry,Braun:1988qv,Chernyak:1990ag,Brodsky:1997de,Ball:1998je,Ball:2006wn} within the framework of heavy quark effective theory (HQET)~\cite{Grinstein:1990mj,Eichten:1989zv,Falk:1990yz}, especially:
\begin{itemize}

\item In Refs.~\cite{Yang:2020zrh,Yang:2021lce} we have systematically investigated the $P$-wave charmed and bottom baryons of the $SU(3)$ flavor $\mathbf{6}_F$, and studied their $S$- and $D$-wave decays into the ground-state heavy baryons and light pseudoscalar/vector mesons.

\item In Ref.~\cite{Chen:2017sci} we have investigated the $P$-wave charmed baryons of the $SU(3)$ flavor $\mathbf{\bar 3}_F$, and studied their $S$-wave decays into the ground-state charmed baryons and light pseudoscalar/vector mesons.

\end{itemize}
In the present study we will further investigate the $P$-wave charmed and bottom baryons of the $SU(3)$ flavor $\mathbf{\bar 3}_F$, and study their $S$- and $D$-wave decays into the ground-state heavy baryons and light pseudoscalar/vector mesons. We shall concentrate on the $\Xi_b(6100)$ recently observed by CMS~\cite{CMS:2021rvl}, and study its mass and decay properties. We refer to Refs.~\cite{Arifi:2021orx,He:2021xrh,Kim:2021ywp,Hazra:2021lpa,Chen:2022fye,Polyakov:2022eub,Kakadiya:2022zvy} for relevant quark model calculations.

Previously in Ref.~\cite{Chen:2017sci} we found that the $\Lambda_c(2595)$, $\Lambda_c(2625)$, $\Xi_c(2790)$, and $\Xi_c(2815)$ can be well explained as the $P$-wave charmed baryons of the $SU(3)$ flavor $\mathbf{\bar 3}_F$, which complete two flavor $\mathbf{\bar 3}_F$ multiplets of $J^P=1/2^-$ and $3/2^-$. In this study we shall find that the $\Lambda_b(5912)$, $\Lambda_b(5920)$, and $\Xi_b(6100)$, as the bottom partner states of the $\Lambda_c(2595)$, $\Lambda_c(2625)$ and $\Xi_c(2815)$, can be well explained as the $P$-wave bottom baryons of the $SU(3)$ flavor $\mathbf{\bar 3}_F$. Hence, there is a $P$-wave bottom baryon still missing, which is the bottom partner state of the $\Xi_c(2790)$. We shall also study its mass and decay properties.

This paper is organized as follows. In Sec.~\ref{sec:sumrule} we briefly introduce our notations, and apply the QCD sum rule method to calculate the mass of $\Xi_b(6100)$ as a $P$-wave bottom baryon of the $SU(3)$ flavor $\mathbf{\bar 3}_F$. The obtained parameters are further used to study its decay properties through the light-cone sum rule method in Sec.~\ref{sec:decay}. In Sec.~\ref{sec:summary} we discuss the results and conclude this paper.

%
\section{Mass analyses from QCD sum rules}
\label{sec:sumrule}
%

\begin{figure*}[hbtp]
\begin{center}
\scalebox{1}{\includegraphics{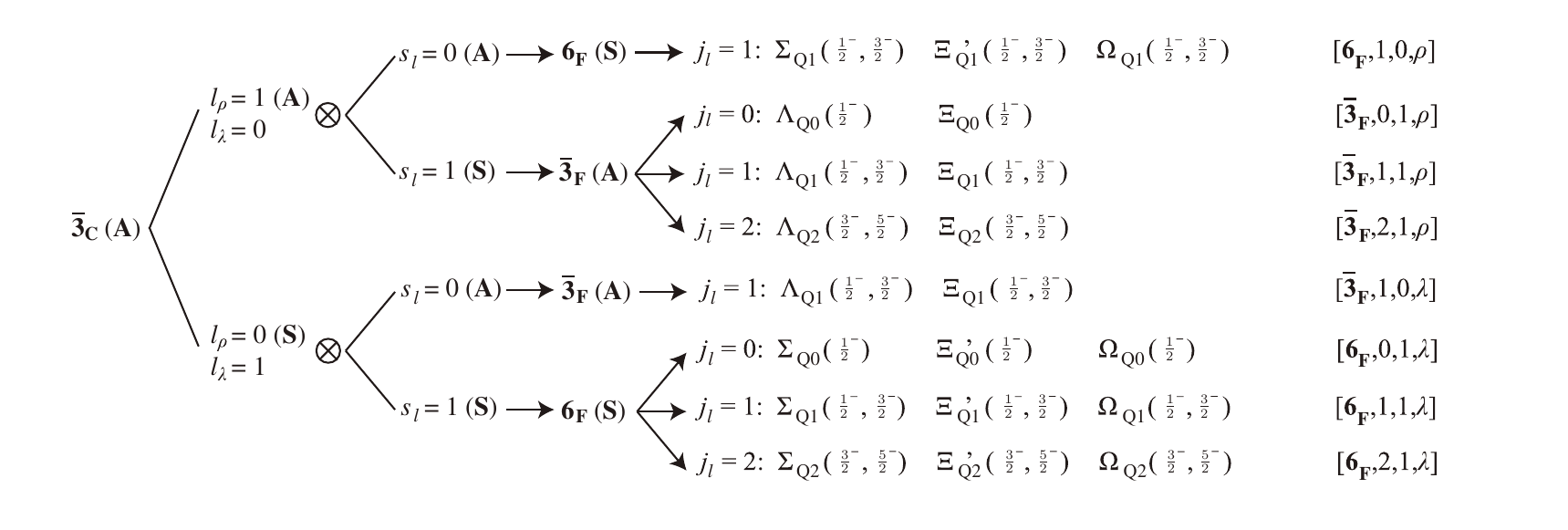}}
\end{center}
\caption{$P$-wave heavy baryons belonging to the $SU(3)$ flavor $\mathbf{\bar 3}_F$ representation.}
\label{fig:pwave}
\end{figure*}

A bottom baryon consists of a heavy bottom quark and two light up/down/strange quarks. It has a rich internal structure with the color, flavor, spin, and orbital degrees of freedom. Especially, the orbital excitation of a $P$-wave bottom baryon can be between the bottom quark and the light quarks, or it can also be inside the two light quarks. We call the former $\lambda$-mode excitation with $l_{\lambda}=1$ and $l_{\rho}=0$, and the latter $\rho$-mode excitation with $l_{\lambda}=0$ and $l_{\rho}=1$.

When investigating the bottom baryon, we need to pay attention to the two light up/down/strange quarks:
\begin{itemize}

\item Their color structure is antisymmetric ($\mathbf{\bar 3}_C$).

\item Their flavor structure is either symmetric ($\mathbf{6}_F$) or antisymmetric ($\mathbf{\bar 3}_F$).

\item Their spin structure is either symmetric ($s_l = 1$) or antisymmetric ($s_l = 0$).

\item Their orbital structure is either symmetric ($\lambda$-mode) or antisymmetric ($\rho$-mode).
\end{itemize}
Applying the Pauli principle to the two light quarks, we can categorize the $P$-wave bottom baryons into eight multiplets. We denote them as $[F({\rm flavor}), j_l, s_l, \rho/\lambda]$, where $j_l = l_\lambda \otimes l_\rho \otimes s_l$ is the total angular momentum of the light components. Every multiplet contains one or two baryons with the total angular momenta $j = j_l \otimes s_b = |j_l \pm 1/2|$. Within the QCD sum rule method we can construct their corresponding interpolating currents $J^{\alpha_1\cdots\alpha_{j-1/2}}_{j,P,F,j_l,s_l,\rho/\lambda}$, which couple to the bottom baryons belonging to the $[F, j_l, s_l, \rho/\lambda]$ multiplet through:
\begin{eqnarray}
\nonumber && \langle 0| J^{\alpha_1\cdots\alpha_{j-1/2}}_{j,P,F,j_l,s_l,\rho/\lambda} |j,P,F,j_l,s_l,\rho/\lambda \rangle
\\  &=& f_{F,j_l,s_l,\rho/\lambda} u^{\alpha_1\cdots\alpha_{j-1/2}} \, ,
\label{eq:coupling}
\end{eqnarray}
with $f_{F,j_l,s_l,\rho/\lambda}$ the decay constant.

As shown in Fig.~\ref{fig:pwave}, there are four multiplets belonging to the $SU(3)$ flavor $\mathbf{\bar 3}_F$ representation. In this paper we study the $\Xi_b(6100)$ as the $P$-wave bottom baryon of $J^P=3/2^-$ belonging to the $[\mathbf{\bar 3}_F, 1, 1, \rho]$ doublet. Accordingly, we concentrate on this doublet and use the following currents to study it~\cite{Chen:2015kpa}:
\begin{eqnarray}
&& J_{1/2,-,\mathbf{\bar 3_F},1,1,\rho}
\\ \nonumber &=&i\epsilon_{abc}\big([\mathcal{D}_t^\mu q^{aT}]\mathcal{C}\gamma_t^\nu q^b-q^{aT}\mathcal{C}\gamma_t^\nu[\mathcal{D}_t^\mu q^b]\big)\sigma^{\mu\nu}h_v^c,\\
&& J_{3/2,-,\mathbf{\bar 3_F},1,1,\rho}^\alpha
\\ \nonumber &=&i\epsilon_{abc}\big([\mathcal{D}_t^\mu q^{aT}]\mathcal{C}\gamma_t^\nu q^b-q^{aT}\mathcal{C}\gamma_t^\nu[\mathcal{D}_t^\mu q^b]\big)\\ \nonumber
&\times& \big(g_t^{\alpha\mu}\gamma_t^\nu\gamma_5-g_t^{\alpha\nu}\gamma_t^\mu\gamma_5-{1\over3}\gamma_t^\alpha\gamma_t^\mu\gamma_t^\nu\gamma_5+{1\over3}\gamma_t^\alpha\gamma_t^\nu\gamma_t^\mu\gamma_5\big)h_v^c \, ,
\end{eqnarray}
where $a \cdots c$ are color indices, $\mathcal{C}$ is the charge-conjugation operator, $\mathcal{D}_t^{\mu}=\mathcal{D}^{\mu}-v\cdot \mathcal{D} v^{\mu}$, $\gamma_t^{\nu}=\gamma^{\nu}-v\!\!\!\slash v^{\nu}$, and $g_t^{\rho\beta}=g^{\rho\beta}-v^{\rho} v^{\beta}$. The covariant derivative operator has been explicitly added to these currents, and we also refer to Ref.~\cite{Dong:2022otb} where we use this covariant derivative operator to construct the $D$-wave fully-strange tetraquark currents with the exotic quantum number $J^{PC} = 4^{+-}$.

We employ QCD sum rules to investigate these two currents and calculate the mass of $\Xi_b(6100)$ through:
\begin{eqnarray}
m_{\Xi_b}&=& m_b + \bar{\Lambda}_{\Xi_b} + \delta m_{\Xi_b},
\\
\delta m_{\Xi_b}&=&-{1\over 4m_b}(K_{\Xi_b}+d_{j,j_l}C_{mag}\Sigma_{\Xi_b})\, ,
\end{eqnarray}
where $m_b$ is the bottom quark mass, $\bar{\Lambda}_{\Xi_b}$ is the sum rule result at the leading order, and $\delta m_{\Xi_b}$ is the result at the $\mathcal{O}(1/m_Q)$ order. 

The obtained results are summarized in Table~\ref{tab:pwavebparameter}, where we use three criteria to restrict the two free parameters: the threshold value $\omega_c$ and the Borel mass $T$. The first criterion requires that the higher-order corrections being less than $30\%$, the second criterion requires that the pole contribution being larger than $20\%$, and the third criterion requires a weak dependence of the mass $m_{\Xi_b}$ on these two free parameters. We refer to Refs.~\cite{Chen:2015kpa,Mao:2015gya} for detailed discussions, where all the QCD sum rule calculations have been done for the $P$-wave charmed and bottom baryons. In the calculations we work at the renormalization scale 2~GeV, and use the following QCD parameters~\cite{pdg,Yang:1993bp,Hwang:1994vp,Ovchinnikov:1988gk,Jamin:2002ev,Ioffe:2002be,Gimenez:2005nt,Colangelo:1998ga}:
\begin{eqnarray}
\nonumber && \langle \bar qq \rangle = - (0.24 \pm 0.01 \mbox{ GeV})^3 \, ,
\\ \nonumber && \langle \bar ss \rangle = (0.8\pm 0.1)\times \langle\bar qq \rangle \, ,
\\ && \langle g_s \bar q \sigma G q \rangle = M_0^2 \times \langle \bar qq \rangle\, ,
\label{eq:condensates}
\\ \nonumber && \langle g_s \bar s \sigma G s \rangle = M_0^2 \times \langle \bar ss \rangle\, ,
\\ \nonumber && M_0^2= 0.8 \mbox{ GeV}^2\, ,
\\ \nonumber && \langle g_s^2GG\rangle =(0.48\pm 0.14) \mbox{ GeV}^4\, ,
\\ \nonumber && m_b = 4.66 \pm 0.03 \mbox{ GeV}\, ,
\\ \nonumber && m_s =95 ^{+9}_{-3}  \mbox{ MeV}\, .
\end{eqnarray}
There are considerable uncertainties in our results for the absolute value of the mass. This allows us to fine-tune some of the parameters in order to get a better description of the $\Xi_b(6100)$. However, the mass difference within the same doublet is produced quite well with much less uncertainties. Moreover, the decay constant $f_{F,j_l,s_l,\rho/\lambda}$ is an important input parameter, which will be used to study the decay properties of $\Xi_b(6100)$ in the next section.

\begin{table*}[hbtp]
\begin{center}
\renewcommand{\arraystretch}{1.5}
\caption{Parameters of the $P$-wave bottom baryons belonging to the $[\mathbf{\bar 3}_F, 1, 1, \rho]$ and $[\mathbf{\bar 3}_F,1,0,\lambda]$ doublets, calculated using the method of QCD sum rules within the framework of heavy quark effective theory. Decay constants in the last column satisfy $f_{\Xi_b^0} = f_{\Xi_b^-}$.}
\begin{tabular}{ c|c | c | c | c | c c | c | c}
\hline\hline
\multirow{2}{*}{Multipluts}&\multirow{2}{*}{~~B~~} & $\omega_c$ & ~~~Working region~~~ & ~~~~~~~$\overline{\Lambda}$~~~~~~~ & ~~~Baryon~~~ & ~~~~Mass~~~~~ & ~Difference~ & Decay constant
\\                                               & & (GeV)      & (GeV)                & (GeV)                              & ($j^P$)       & (GeV)      & (MeV)        & (GeV$^{4}$)
\\ \hline\hline
 \multirow{4}{*}{$[\mathbf{\bar 3}_F,1,1,\rho]$}&\multirow{2}{*}{$\Lambda_b$} & \multirow{2}{*}{1.63} & \multirow{2}{*}{$0.27< T < 0.31$} & \multirow{2}{*}{$1.20 \pm 0.10$} & $\Lambda_b(1/2^-)$ & $5.92^{+0.13}_{-0.10}$ & \multirow{2}{*}{$10 \pm 4$} & $0.060 \pm 0.014~(\Lambda^0_b(1/2^-))$
\\ \cline{6-7}\cline{9-9}
 & & & & & $\Lambda_b(3/2^-)$ & $5.93^{+0.13}_{-0.10}$ & &$0.028 \pm 0.007~(\Lambda^0_b(3/2^-))$
\\ \cline{2-9}
 &\multirow{2}{*}{$\Xi_b$} & \multirow{2}{*}{1.83} & \multirow{2}{*}{$0.27< T < 0.32$} & \multirow{2}{*}{$1.37 \pm 0.10$} & $\Xi_b(1/2^-)$ & $6.08^{+0.13}_{-0.11}$ & \multirow{2}{*}{$9 \pm 3$} & $0.083 \pm 0.020~(\Xi_b^-(1/2^-))$
\\ \cline{6-7}\cline{9-9}
 & & & & & $\Xi_b(3/2^-)$ & $6.09^{+0.12}_{-0.11}$ & &$0.039 \pm 0.009~(\Xi_b^-(3/2^-))$
\\ \hline\hline
 \multirow{4}{*}{$[\mathbf{\bar 3}_F,1,0,\lambda]$}&\multirow{2}{*}{ $\Lambda_b$} &\multirow{2}{*}{ $1.70$ }& \multirow{2}{*}{$0.32< T < 0.33$} &\multirow{2}{*}{ $1.13\pm 0.10$} & $\Lambda_b(1/2^-)$ & $5.91 ^{+0.11}_{-0.11}$ &\multirow{2}{*}{ $6\pm2$}& $0.031 \pm 0.007~(\Lambda^0_b(1/2^-))$
\\ \cline{6-7}\cline{9-9}
& & & & &$\Lambda_b(3/2^-)$ & $5.92^{+0.11}_{-0.11}$ & &$0.015 \pm 0.003~(\Lambda^0_b(3/2^-))$
\\ \cline{2-9}
                                                & \multirow{2}{*}{$\Xi_b$} & \multirow{2}{*}{$1.83$} & \multirow{2}{*}{$0.33< T < 0.35$} & \multirow{2}{*}{$1.28 \pm 0.09$} & $\Xi_b(1/2^-)$ & $6.09^{+0.10}_{-0.10}$ & \multirow{2}{*}{$5\pm 2$} & $0.044 \pm 0.009~(\Xi^-_b(1/2^-))$
                                                 \\ \cline{6-7}\cline{9-9}
& & & & & $\Xi_b(3/2^-)$ & $6.10 ^{+0.10}_{-0.10}$ & &$0.021 \pm 0.004~(\Xi_b^-(3/2^-))$
\\ \hline\hline
\end{tabular}
\label{tab:pwavebparameter}
\end{center}
\end{table*}

Besides the $[\mathbf{\bar 3}_F, 1, 1, \rho]$ doublet, we have also investigated the $[\mathbf{\bar 3}_F,1,0,\lambda]$ doublet through the following interpolating currents:
\begin{eqnarray}
\nonumber J_{1/2,-,\mathbf{\bar 3_F},1,0,\lambda}&=&i\epsilon_{abc}\big([\mathcal{D}_t^\mu q^{aT}]\mathcal{C}\gamma_5 q^b+q^{aT}\mathcal{C}\gamma_5[\mathcal{D}_t^\mu q^b]\big)
\\
&\times& \gamma_t^\mu\gamma_5h_v^c,\\
\nonumber J_{3/2,-,\mathbf{\bar 3_F},1,0,\lambda}^\alpha&=&i\epsilon_{abc}\big([\mathcal{D}_t^\mu q^{aT}]\mathcal{C}\gamma_5 q^b+q^{aT}\mathcal{C}\gamma_5[\mathcal{D}_t^\mu q^b]\big)
\\
&\times& \big(g_t^{\alpha\mu}-{1\over3}\gamma_t^\alpha\gamma_t^\mu\big)h_v^c \, .
\end{eqnarray}
The obtained results are also summarized in Table~\ref{tab:pwavebparameter}. We find that the $\rho$-mode doublet $[\mathbf{\bar 3}_F,1,1,\rho]$ is lower than the $\lambda$-mode doublet $[\mathbf{\bar 3}_F,1,0,\lambda]$. This behavior is consistent with our previous QCD sum rule results for their corresponding doublets of the $SU(3)$ flavor $\mathbf{6}_F$~\cite{Yang:2021lce,Yang:2020zrh,Yang:2020zjl}, but in contrast to the quark model expectation~\cite{Yoshida:2015tia,Nagahiro:2016nsx}. However, this is possible simply because the mass difference between different multiplets have considerable uncertainties within our QCD sum rule framework, so our results for the mass spectra can not distinguish the $[\mathbf{\bar 3}_F,1,1,\rho]$ and $[\mathbf{\bar 3}_F,1,0,\lambda]$ doublets. Therefore, we move on to investigate their decay properties in the next section.

\section{Decay analyses from light-cone sum rules}
\label{sec:decay}

In this section we study the decay properties of $\Xi_b(6100)$ as the $P$-wave bottom baryon of $J^P=3/2^-$ belonging to the $[\mathbf{\bar 3}_F, 1, 1, \rho]$ doublet. We shall apply the method of light-cone sum rules to investigate all the bottom baryons from this doublet, and study their $S$- and $D$-wave decays into the ground-state bottom baryons and light pseudoscalar/vector mesons, including:
\begin{widetext}
\begin{eqnarray}
 &(a1)& {\bf \Gamma\Big[} \Lambda_b[1/2^-] \rightarrow \Lambda_b + \pi {\Big]}
=  {\bf \Gamma\Big[}\Lambda_b^0[1/2^-] \rightarrow \Lambda_b^0 +\pi^0 {\Big]} \, ,
\label{eq:couple1}
\\ &(a2)& {\bf \Gamma\Big[} \Lambda_b[1/2^-] \rightarrow \Sigma_b + \pi {\Big]}
=3\times {\bf \Gamma\Big[} \Lambda_b^0[1/2^-] \rightarrow \Sigma_b^++\pi^-\to\Lambda_b^0+\pi^++\pi^- {\Big]} \, ,
\\ &(a3)& {\bf \Gamma\Big[} \Lambda_b[1/2^-] \rightarrow \Sigma_b^* + \pi\rightarrow\Lambda_b+\pi+\pi {\Big]}
= 3 \times {\bf \Gamma \Big[}\Lambda_b^0[1/2^-] \rightarrow \Sigma_b^{*+}+\pi^-\rightarrow \Lambda_b^0+\pi^++\pi^-{\Big ]} \, ,
\\ &(a4)& {\bf\Gamma\Big[} \Lambda_b[1/2^-] \rightarrow \Lambda_b + \rho \rightarrow\Lambda_b+\pi+\pi{\Big ]}
= {\bf \Gamma\Big[} \Lambda_b^0[1/2^-] \rightarrow \Lambda_b^0 +\pi^++ \pi^- {\Big ]} \, ,
\\ &(a5)& { \bf\Gamma\Big[}\Lambda_b[1/2^-] \rightarrow \Sigma_b + \rho\rightarrow\Sigma_b+\pi+\pi{\Big ]}
= 3 \times { \bf\Gamma\Big[}\Lambda_b^0[1/2^-] \rightarrow \Sigma_b^0 +\pi^++ \pi^-{\Big ]} \, ,
\\ &(a6)&{\bf \Gamma\Big[}\Lambda_b[1/2^-] \rightarrow \Sigma_b^* + \rho\rightarrow\Sigma_b^*+\pi+\pi{\Big ]}
= 3 \times { \bf\Gamma\Big[}\Lambda_b^0[1/2^-] \rightarrow \Sigma_b^{*0} +\pi^++ \pi^-{\Big ]} \, ,
\\ &(b1)& {\bf \Gamma\Big[}\Lambda_b[3/2^-] \rightarrow \Lambda_b + \pi{\Big ]}
= {\bf \Gamma\Big[}\Lambda_b^0[3/2^-] \rightarrow \Lambda_b^0 +\pi^0{\Big ]} \, ,
\\ &(b2)&{\bf \Gamma\Big[}\Lambda_b[3/2^-] \rightarrow \Sigma_b + \pi\rightarrow\Lambda_b+\pi+\pi{\Big ]}
= 3 \times {\bf \Gamma\Big[}\Lambda_b^0[3/2^-] \rightarrow \Sigma_b^+ +\pi^-\rightarrow\Lambda_b^0+\pi^++\pi^-{\Big ]} \, ,
\\ &(b3)&{\bf \Gamma\Big[}\Lambda_b[3/2^-] \rightarrow \Sigma_b^* + \pi\rightarrow\Lambda_b+\pi+\pi{\Big ]}
= 3 \times {\bf \Gamma\Big[}\Lambda_b^0[3/2^-] \rightarrow \Sigma_b^{*+} +\pi^-\rightarrow\Lambda_b^0+\pi^+\pi^-{\Big ]} \, ,
\\ &(b4)&{\bf \Gamma\Big[} \Lambda_b[3/2^-] \rightarrow \Lambda_b + \rho \rightarrow\Lambda_b+\pi+\pi{\Big ]}
= { \bf\Gamma\Big[}\Lambda_b^0[3/2^-] \rightarrow \Lambda_b^0 +\pi^++ \pi^- {\Big ]} \, ,
\\ &(b5)& { \bf\Gamma\Big[}\Lambda_b[3/2^-] \rightarrow \Sigma_b + \rho\rightarrow\Sigma_b+\pi+\pi{\Big ]}
= 3 \times { \bf\Gamma\Big[}\Lambda_b^0[3/2^-] \rightarrow \Sigma_b^0 +\pi^++ \pi^-{\Big ]} \, ,
\\&(b6)& { \bf\Gamma\Big[}\Lambda_b[3/2^-] \rightarrow \Sigma^* + \rho\rightarrow\Sigma_b^*+\pi+\pi {\Big ]}
= 3 \times {\bf \Gamma\Big[}\Lambda_b^0[3/2^-] \rightarrow \Sigma_b^{*0} + \pi^++\pi^- {\Big ]} \, ,
\\ &(c1)& {\bf \Gamma\Big[}\Xi_b[1/2^-] \rightarrow \Lambda_b + \bar K{\Big ]}
= {\bf \Gamma\Big[}\Xi_b^-[1/2^-] \rightarrow\Lambda_b^0 +K^-{\Big ]} \, ,
\\ &(c2)&{\bf \Gamma\Big[}\Xi_b[1/2^-] \rightarrow\Xi_b + \pi{\Big ]}
= {3\over2} \times {\bf \Gamma\Big[}\Xi_b^-[1/2^-] \rightarrow \Xi_b^0 +\pi^-{\Big ]} \, ,
\\ &(c3)&{\bf \Gamma\Big[}\Xi_b[1/2^-] \rightarrow\Sigma_b + \bar K{\Big ]}
= 3 \times {\bf \Gamma\Big[}\Xi_b^-[1/2^-] \rightarrow \Sigma_b^0 +K^-{\Big ]} \, ,
\\ &(c4)& { \bf\Gamma\Big[}\Xi_b[1/2^-] \rightarrow \Xi_b^{\prime}+\pi {\Big]}
= {3\over2}\times{\bf \Gamma\Big[}\Xi_b^-[1/2^-]\rightarrow\Xi_b^{\prime0}+\pi^-{\Big]} \, ,
\\ &(c5)& {\bf \Gamma\Big[}\Xi_b[1/2^-] \rightarrow \Sigma_b^* + K{\Big ]}
= 3 \times {\bf \Gamma\Big[}\Xi_b^-[1/2^-] \rightarrow \Sigma_b^{*0} + \bar K^-{\Big ]} \, ,
\\ &(c6)& {\bf \Gamma\Big[}\Xi_b[1/2^-] \rightarrow \Xi_b^* + \pi\rightarrow \Xi_b+\pi+\pi {\Big ]}
= {9 \over 2} \times {\bf \Gamma\Big[}\Xi_b^{*-}[1/2^-] \rightarrow\Xi_b^{*0} + \pi^-\rightarrow\Xi_b^0+\pi^0+\pi^-{\Big]} \, ,
\\ &(c7)& {\bf \Gamma\Big[}\Xi_b[1/2^-] \rightarrow \Lambda_b + \bar{K}^*\rightarrow\Lambda_b+\bar K+\pi{\Big ]}
=3\times  {\bf \Gamma\Big[}\Xi_b^-[1/2^-] \rightarrow \Lambda_b^0 +  K^- +\pi^0{\Big ]} \, ,
\\ &(c8)& {\bf \Gamma\Big[}\Xi_b[1/2^-] \rightarrow\Xi_b + \rho\rightarrow\Xi_b+\pi+\pi{\Big ]}
= {3\over2} \times {\bf \Gamma\Big[}\Xi_b^-[1/2^-] \rightarrow \Xi_b^0 + \pi^0+\pi^-{\Big ]} \, ,
\\ &(c9)& {\bf \Gamma\Big[}\Xi_b[1/2^-] \rightarrow \Sigma_b^* + \bar K^*\rightarrow\Sigma_b^{*0}+\bar K+\pi{\Big ]}
= 9 \times {\bf \Gamma\Big[}\Xi_b^-[1/2^-] \rightarrow \Sigma_b^{*0} + K^-+ \pi^0{\Big ]} \, ,
\\ &(c10)& {\bf \Gamma\Big[}\Xi_b[1/2^-] \rightarrow \Xi_b^{*}+ \rho\rightarrow\Xi_b^{*}+\pi+\pi {\Big ]}
= {3\over2} \times {\bf \Gamma\Big[}\Xi_b^-[1/2^-] \rightarrow \Xi_b^{*0} + \pi^0+\pi^- {\Big ]} \, .
\\ &(d1)& {\bf \Gamma\Big[}\Xi_b[3/2^-] \rightarrow \Lambda_b + \bar K{\Big ]}
= {\bf \Gamma\Big[}\Xi_b^-[3/2^-] \rightarrow\Lambda_b^0 +K^-{\Big ]} \, ,
\\ &(d2)&{\bf \Gamma\Big[}\Xi_b[3/2^-] \rightarrow\Xi_b + \pi{\Big ]}
= {3\over2} \times {\bf \Gamma\Big[}\Xi_b^-[3/2^-] \rightarrow \Xi_b^0 +\pi^-{\Big ]} \, ,
\\ &(d3)&{\bf \Gamma\Big[}\Xi_b[3/2^-] \rightarrow \Sigma_b + \bar K{\Big ]}
= 3 \times {\bf \Gamma\Big[}\Xi_b^-[3/2^-] \rightarrow \Sigma_b^0 +K^-{\Big ]} \, ,
\\ &(d4)& { \bf\Gamma\Big[}\Xi_b[3/2^-] \rightarrow \Xi_b^{\prime}+\pi {\Big]}
= {3\over2}\times{\bf \Gamma\Big[}\Xi_b^-[3/2^-]\rightarrow\Xi_b^{\prime0}+\pi^-{\Big]} \, ,
\\ &(d5)& {\bf \Gamma\Big[}\Xi_b[3/2^-] \rightarrow \Sigma_b^* + \bar K{\Big ]}
= 3 \times {\bf \Gamma\Big[}\Xi_b^-[3/2^-] \rightarrow \Sigma_b^{*0} + K^-{\Big ]} \, ,
\\ &(d6)& {\bf \Gamma\Big[}\Xi_b[3/2^-] \rightarrow \Xi_b^* + \pi\rightarrow \Xi_b+\pi+\pi {\Big ]}
= {9 \over 2} \times {\bf \Gamma\Big[}\Xi_b^{*-}[3/2^-] \rightarrow\Xi_b^{*0} + \pi^-\rightarrow\Xi_b^0+\pi^0+\pi^-{\Big]} \, ,
\\ &(d7)& {\bf \Gamma\Big[}\Xi_b[3/2^-] \rightarrow \Lambda_b + \bar{K}^*\rightarrow\Lambda_b+\bar K+\pi{\Big ]}
=3\times  {\bf \Gamma\Big[}\Xi_b^-[3/2^-] \rightarrow \Lambda_b^0 +  K^- +\pi^0{\Big ]} \, ,
\\ &(d8)& {\bf \Gamma\Big[}\Xi_b[3/2^-] \rightarrow \Xi_b + \rho\rightarrow\Xi_b+\pi+\pi{\Big ]}
= {3\over2} \times {\bf \Gamma\Big[}\Xi_b^-[3/2^-] \rightarrow \Xi_b^0 + \pi^0+\pi^-{\Big ]} \, ,
\\ &(d9)& {\bf \Gamma\Big[}\Xi_b[3/2^-] \rightarrow \Sigma_b^* + \bar K^*\rightarrow\Sigma_b^{*0}+\bar K+\pi{\Big ]}
= 9 \times {\bf \Gamma\Big[}\Xi_b^-[3/2^-] \rightarrow \Sigma_b^{*0} + K^-+ \pi^0{\Big ]} \, ,
\\ &(d10)& {\bf \Gamma\Big[} \Xi_b[3/2^-] \rightarrow \Xi_b^{*}+ \rho\rightarrow\Xi_b^{*}+\pi+\pi {\Big ]}
= {3\over2} \times {\bf \Gamma\Big[}\Xi_b^-[3/2^-] \rightarrow \Xi_b^{*0} + \pi^0+\pi^- {\Big ]} \, .
\label{eq:couple28}
\end{eqnarray}
We shall calculate their partial decay widths through the following Lagrangians:
\begin{eqnarray}
&&\mathcal{L}^S_{X_b({1/2}^-) \rightarrow Y_b({1/2}^+) \mathcal P} = g {\bar X_b}(1/2^-) Y_b(1/2^+) \mathcal P \, ,
\\ &&\mathcal{L}^S_{X_b({3/2}^-) \rightarrow Y_b({3/2}^+) \mathcal P} = g {\bar X_{b\mu}}(3/2^-)Y_b^{\mu}(3/2^+) \mathcal P \, ,
\\ &&\mathcal{L}^S_{X_b({1/2}^-) \rightarrow Y_b({1/2}^+) V} = g {\bar X_b}(1/2^-) \gamma_\mu \gamma_5 Y_b(1/2^+) V^\mu \, ,
\\ &&\mathcal{L}^S_{X_b({1/2}^-) \rightarrow Y_b({3/2}^+) V} = g {\bar X_{b}}(1/2^-) Y_{b}^{\mu}(3/2^+) V_\mu \, ,
\\ &&\mathcal{L}^S_{X_b({3/2}^-) \rightarrow Y_b({1/2}^+) V} = g {\bar X_{b}^{\mu}}(3/2^-) Y_{b}(1/2^+) V_\mu \, ,
\\ &&\mathcal{L}^S_{X_b({3/2}^-) \rightarrow Y_b({3/2}^+) V} = g {\bar X_b}^{\nu}(3/2^-) \gamma_\mu \gamma_5 Y_{b\nu}(3/2^+) V^\mu \, .
\\ && \mathcal{L}^D_{X_b({1/2}^-) \rightarrow Y_b({3/2}^+) \mathcal P} = g {\bar X_b}(1/2^-) \gamma_\mu \gamma_5 Y_{b\nu}(3/2^+) \partial^{\mu} \partial^{\nu}\mathcal P \, ,
\\ && \mathcal{L}^D_{X_b({3/2}^-) \rightarrow Y_b({1/2}^+) \mathcal P} = g {\bar X_{b\mu}}(3/2^-) \gamma_\nu \gamma_5 Y_{b}(1/2^+) \partial^{\mu} \partial^{\nu}\mathcal P \, ,
\\ && \mathcal{L}^D_{X_b({3/2}^-) \rightarrow Y_b({3/2}^+) \mathcal P} = g {\bar X_{b\mu}}(3/2^-) Y_{b\nu}(3/2^+) \partial^{\mu} \partial^{\nu}\mathcal P \, .
\end{eqnarray}
\end{widetext}
In the above expression, the superscripts $S$ and $D$ denote the $S$- and $D$-wave decays, respectively; the fields $X_b^{(\mu)}$, $Y_b^{(\mu)}$, $\mathcal{P}$, and $V^\mu$ denote the $P$-wave bottom baryons, ground-state bottom baryons, light pseudoscalar mesons, and light vector mesons, respectively.

We use the $\Xi_b^-({3/2}^-)$ from the $[\mathbf{\bar 3}_F, 1, 1, \rho]$ doublet as an example, and study its $S$-wave decays into the ground-state bottom baryon $\Xi_b^{*0}$ and the light pseudoscalar meson $\pi^-$. We note that the two-body decay is insufficient due to the $\Xi_b^*\pi$ threshold being close to the mass of $\Xi_b(6100)$, so the sequential three-body decay process $\Xi_b(3/2^-)\to \Xi_b^*(3/2^+)\pi\to \Xi_b(1/2^+)\pi\pi$ is crucial for understanding the decay behavior of $\Xi_b(6100)$. To do this we investigate the three-point correlation function:
\begin{eqnarray}
&& \Pi^\alpha(\omega, \omega^\prime)
\\ \nonumber &=& \int d^4 x~e^{-i k \cdot x}~\langle 0 | J^\alpha_{3/2,-,\Xi_b^-,1,1,\rho}(0) \bar J_{\Xi_b^{*0}}(x) | \pi^-(q) \rangle
\\ \nonumber &=& {1+v\!\!\!\slash\over2} G^\alpha_{\Xi_b^-[{3\over2}^-] \rightarrow   \Xi_b^{*0}\pi^-} (\omega, \omega^\prime) \, ,
\end{eqnarray}
where $k^\prime = k + q$, $\omega = v \cdot k$, and $\omega^\prime = v \cdot k^\prime$.

At the hadron level we write
$G^{\alpha\beta}_{\Xi_b^-[{3\over2}^-] \rightarrow \Xi_b^{*0}\pi^-}$ as:
\begin{eqnarray}
&&G^{\alpha\beta}_{\Xi_b^-[{3\over2}^-] \rightarrow \Xi_b^{*0}\pi^-} (\omega, \omega^\prime)\\ \nonumber
&=& g_{\Xi_b^-[{3\over2}^-] \rightarrow \Xi_b^{*0}\pi^-} { g^{\alpha\beta} f_{\Xi_b^-[{3\over2}^-]} f_{\Xi_b^{*0}} \over (\bar \Lambda_{\Xi_b^-[{3\over2}^-]} - \omega^\prime) (\bar \Lambda_{\Xi_b^{*0}} - \omega)} + \cdots \, ,
\label{G0C}
\end{eqnarray}
where $\cdots$ contains other possible amplitudes.

At the quark-gluon level we calculate $G^\alpha_{\Xi_b^-[{3\over2}^-] \rightarrow \Xi_b^{*0}\pi^-}$ using the method of operator product expansion (OPE):
\begin{widetext}
\begin{eqnarray}
\label{eq:sumrule}
&& G^{\alpha\beta}_{\Xi_b^-[{3\over2}^-] \rightarrow \Xi_b^{*0}\pi^-} (\omega, \omega^\prime)
\\ \nonumber &=& \int_0^\infty dt \int_0^1 du e^{i (1-u) \omega^\prime t} e^{i u \omega t} \times 4 \times \Big (
\frac{f_\pi m_s t^2 v\cdot q}{1152}\langle\bar s s\rangle\phi_{4;\pi}(u)-\frac{f_\pi m_s v\cdot q}{72}\langle\bar s s\rangle\phi_{2;\pi}(u)
\\ \nonumber &&
-\frac{f_\pi m_\pi^2 m_s v\cdot q}{36(m_u+m_d)\pi^2t^2}\phi^\sigma_{3;\pi}(u)
-\frac{5if_\pi m_sut^3(v\cdot q)^2}{13824}\langle\bar s s\rangle\phi_{4;\pi}(u)
-\frac{5if_\pi m_s ut(v\cdot q)^2}{864}\langle\bar s s\rangle\phi_{2;\pi}(u)
\\ \nonumber &&
+\frac{5if_\pi m_\pi^2m_s u(v\cdot q)^2}{432(m_u+m_d)\pi^2 t}\phi^\sigma_{3;\pi}(u)
-\frac{f_\pi m_s}{72v\cdot q}\langle\bar s s\rangle\psi_{4;\pi}(u)
-\frac{f_\pi m_\pi^2 t^2 v\cdot q}{1728(m_u+m_d)}\langle g_s\bar s \sigma G s\rangle\phi^\sigma_{3;\pi}(u)
\\ \nonumber &&
-\frac{f_\pi m_\pi^2v\cdot q}{108(m_u+m_d)}\langle\bar s s\rangle\phi^\sigma_{3;\pi}(u)
-\frac{f_\pi v\cdot q}{48\pi^2t^2}\phi_{4;\pi}(u)
-\frac{f_\pi v\cdot q}{3\pi t^4}\phi_{2;\pi}(u)
\\ \nonumber &&
+\frac{5if_\pi m_\pi^2ut^3(v\cdot q)^2}{20736(m_u+m_d)}\langle g_s \bar s\sigma G s\rangle\phi^\sigma_{3;\pi}(u)+\frac{5if_\pi m_\pi^2 ut(v\cdot q)^2}{1296(m_u+m_d)}\langle\bar s s\rangle\phi^\sigma_{3;\pi}(u)-\frac{5if_\pi u(v\cdot q)^2}{576\pi^2 t}\phi_{4;\pi}(u)
\\ \nonumber &&
-\frac{5if_\pi u(v\cdot q)^2}{36\pi^2 t^3}\phi_{2;\pi}(u)-\frac{f_\pi}{3\pi^2t^4v\cdot q}\psi_{4;\pi}(u)\Big )\times g^{\alpha\beta}
\\ \nonumber &-&
\int_0^\infty dt \int_0^1 du \int \mathcal{D} \underline{\alpha} e^{i \omega^{\prime} t(\alpha_2 + u \alpha_3)} e^{i \omega t(1 - \alpha_2 - u \alpha_3)} \times{1\over2}\times \Big (-\frac{5if_\pi\alpha_3 u^2(v\cdot q)^2}{72\pi^2t}\Phi_{4;\pi}(\underline{\alpha})-\frac{f_\pi u}{36\pi^2t^2}\Phi_{4;\pi}(\underline{\alpha})
\\ \nonumber &&
+\frac{f_\pi u v\cdot q}{12\pi^2 t^2}\widetilde\Phi_{4;\pi}(\underline{\alpha})-\frac{7f_\pi u v\cdot q}{72 \pi^2 t^2}\Psi_{4;\pi}(\underline{\alpha})+\frac{f_\pi u v\cdot q}{24\pi^2 t^2}\widetilde\Psi_{4;\pi}(\underline{\alpha})
\\ \nonumber &&
-\frac{5if_\pi\alpha_2 u(v\cdot q)^2}{72\pi^2 t}\Phi_{4;\pi}(\underline{\alpha})-\frac{5if_\pi\alpha_3u(v\cdot q)^2}{144\pi^2t}\Phi_{4;\pi}(\underline{\alpha})-\frac{5if_\pi\alpha_3u(v\cdot q)^2}{144\pi^2t}\widetilde\Phi_{4;\pi}(\underline{\alpha})
\\ \nonumber &&
+\frac{5if_\pi u(v\cdot q)^2}{72\pi^2t}\Phi_{4;\pi}(\underline{\alpha})-\frac{f_\pi v\cdot q}{24\pi^2t^2}\Phi_{4;\pi}(\underline{\alpha})-\frac{f_\pi v\cdot q}{72\pi^2t^2}\widetilde\Phi_{4;\pi}(\underline{\alpha})
\\ \nonumber &&
+\frac{5f_\pi v\cdot q}{72\pi^2t^2}\Psi_{4;\pi}(\underline{\alpha})-\frac{5f_\pi v\cdot q}{72\pi^2t^2}\widetilde\Psi_{4;\pi}(\underline{\alpha})-\frac{5if_\pi\alpha_2(v\cdot q)^2}{144\pi^2t}\Phi_{4;\pi}(\underline{\alpha})
\\ \nonumber &&
-\frac{5if_\pi\alpha_2(v\cdot q)^2}{144\pi^2t}\widetilde\Phi_{4;\pi} (\underline{\alpha})+\frac{5if_\pi(v\cdot q)^2}{144\pi^2t}\Phi_{4;\pi}(\underline{\alpha})+\frac{5if_\pi(v\cdot q)^2}{144\pi^2t}\widetilde\Phi_{4;\pi}(\underline{\alpha})
\Big ) \times g^{\alpha\beta}\, .
\end{eqnarray}
\end{widetext}
Then we perform the double Borel transformation to both Eq.~(\ref{G0C}) at the hadron level and Eq.~(\ref{eq:sumrule}) at the quark-gluon level:
\begin{widetext}
\begin{eqnarray}
\label{eq:g}
&& g^S_{\Xi_b^-[{3\over2}^-] \rightarrow \Xi_b^{*0}\pi^-} f_{\Xi_b^-[{3\over2}^-]} f_{\Xi_b^{*0}} e^{- {\bar \Lambda_{\Xi_b^-[{3\over2}^-]} \over T_1}} e^{ - {\bar \Lambda_{\Xi_b^{*0}} \over T_2}}
\\ \nonumber &=& 4 \times \Big (-\int_0^{1\over2}du_1\frac{if_\pi}{3\pi^2}T^4f_3({\omega_c\over T})\psi_{4;\pi}(u_1)-\frac{if_\pi}{3\pi^2}T^6f_5({\omega_c\over T})\frac{\partial}{\partial u_0}\phi_{2;\pi}(u_0)|_{u_0={1\over2}}-\frac{5if_\pi}{36\pi^2}T^6f_5({\omega_c\over T})\frac{\partial^2}{\partial u_0^2}u_0\phi_{2;\pi}(u_0)|_{u_0={1\over2}}
\\ \nonumber &&
+\frac{if_\pi}{48\pi^2}T^4f_3({\omega_c\over T})\frac{\partial}{\partial u_0}\phi_{4;\pi}(u_0)|_{u_0={1\over2}}
+\frac{if_\pi m_\pi^2 m_s}{36(m_u+m_d)\pi^2}T^4f_3({\omega_c\over T})\frac{\partial}{\partial u_0}\phi^\sigma_{3;\pi}(u_0)|_{u_0={1\over2}}
\\ \nonumber &&+\frac{5if_\pi}{576\pi^2}T^4f_3({\omega_c\over T})\frac{\partial}{\partial u_0}u_0\phi_{4;\pi}(u_0)|_{u_0={1\over2}}
-\frac{5if_\pi m_\pi^2 m_s}{432(m_u+m_d)\pi^2}T^4f_3({\omega_c\over T})\frac{\partial^2}{\partial u_0^2}u_0\phi^\sigma_{3;\pi}(u_0)|_{u_0={1\over2}}
\\ \nonumber &&
-\frac{i f_\pi m_s}{72}\langle \bar s s\rangle\int_0^{1\over2}du_1\psi_{4;pi}(u_1)
-\frac{if_\pi m_s}{72}\langle\bar s s\rangle T^2f_1({\omega\over T})\frac{\partial}{\partial u_0}\phi_{2;\pi}(u_0)|_{u_0={1\over2}}
\\ \nonumber &&-\frac{if_\pi m_\pi^2}{108(m_u+m_d)}\langle\bar s s\rangle T^2f_1({\omega_c\over T})\frac{\partial}{\partial u_0}\phi^\sigma_{3;\pi}(u_0)|_{u_0={1\over2}}-\frac{5if_\pi m_s}{864}\langle\bar s s\rangle T^2f_1({\omega_c\over T})\frac{\partial^2}{\partial u_0^2}u_0\phi_{2;\pi}(u_0)|_{u_0={1\over2}}
\\ \nonumber &&+\frac{5if_\pi m_\pi^2}{1296(m_u+m_d)}\langle\bar s s\rangle T^2f_1({\omega_c\over T})\frac{\partial^2}{\partial u_0^2}u_0\phi^\sigma_{3;\pi}(u_0)|_{u_0={1\over2}}
+\frac{if_\pi m_s}{1152}\langle\bar s s\rangle\frac{\partial}{\partial u_0}\phi_{4;\pi}(u_0)|_{u_0={1\over2}}
\\ \nonumber &&+\frac{if_\pi m_\pi^2}{1728(m_u+m_d)}\langle g_s\bar s\sigma G s\rangle\frac{\partial}{\partial u_0}\phi^\sigma_{3;\pi}(u_0)|_{u_0={1\over2}}+\frac{5if_\pi m_s}{13824}\langle\bar s s\rangle\frac{\partial^2}{\partial u_0^2}u_0\phi_{4;\pi}(u_0)|_{u_0={1\over2}}
\\ \nonumber &&-\frac{5if_\pi m_\pi^2}{20736(m_u+m_d)}\langle g_s\bar s\sigma G s\rangle\frac{\partial^2}{\partial u_0^2}u_0\phi^\sigma_{3;\pi}(u_0)|_{u_0={1\over2}}\Big )
\\ \nonumber &-&
{1\over2}\times\Big(\frac{if_\pi}{12\pi^2 u_0}T^4f_3({\omega_c\over T})\int_0^{1\over 2}d\alpha_2\int_{{1\over2}-\alpha_2}^{1-\alpha_2}(-\frac{u_0}{3\alpha_3}\frac{\partial}{\partial \alpha_3}\Phi_{4;\pi}(\underline{\alpha})+\frac{u_0}{\alpha_3}\frac{\partial}{\partial \alpha_3}\widetilde\Phi_{4;\pi}(\underline{\alpha})-\frac{7u_0}{6\alpha_3}\frac{\partial}{\partial\alpha_3}\Psi_{4;\pi}(\underline{\alpha})
\\ \nonumber &&
+\frac{u_0}{2\alpha_3}\frac{\partial}{\partial\alpha_3}\widetilde\Psi_{4;\pi}(\underline{\alpha})-\frac{1}{2\alpha_3}\frac{\partial}{\partial\alpha_3}\Phi_{4;\pi}(\underline{\alpha})-\frac{1}{6\alpha_3}\frac{\partial}{\partial\alpha_3}\widetilde\Phi_{4;\pi}(\underline{\alpha})+\frac{5}{6\alpha_3}\frac{\partial}{\partial\alpha_3}\Psi_{4;\pi}(\underline{\alpha})-\frac{5}{6\alpha_3}\frac{\partial}{\partial\alpha_3}\widetilde\Psi_{4;\pi}(\underline{\alpha}))
\\ \nonumber &&
-\frac{5i f_\pi}{72\pi^2u_0^2}T^4f_3({\omega_c\over T}) \int_0^{1 \over 2} d\alpha_2 \int_{{1 \over 2}-\alpha_2}^{1-\alpha_2} d\alpha_3(\frac{-u_0^2}{\alpha_3}\frac{\partial^2}{\partial\alpha_3^2}\alpha_3\Phi_{4;\pi}(\underline{\alpha})-\frac{\alpha_2u_0}{\alpha_3} \frac{\partial^2}{\partial\alpha_3^2}\Phi_{4;\pi}(\underline{\alpha})-\frac{u_0}{2\alpha_3}\frac{\partial^2}{\partial\alpha_3^2}\Phi_{4;\pi}(\underline{\alpha})
\\ \nonumber &&
-\frac{u_0}{2\alpha_3}\frac{\partial^2}{\partial\alpha_3^2}\alpha_3\widetilde\Phi_{4;\pi}(\underline{\alpha})+\frac{u_0}{\alpha_3}\frac{\partial^2}{\partial\alpha_3^2}\Phi_{4;\pi}(\underline{\alpha}) -\frac{\alpha_2}{2\alpha_3}\frac{\partial^2}{\partial\alpha_3^2}\Phi_{4;\pi}(\underline{\alpha})-\frac{\alpha_2}{2\alpha_3}\frac{\partial^2}{\partial\alpha_3}\widetilde\Phi_{4;\pi}(\underline{\alpha})
\\ \nonumber &&
+\frac{1}{2\alpha_3}\frac{\partial^2}{\partial\alpha_3^2}\Phi_{4;\pi}(\underline{\alpha})+\frac{1}{2\alpha_3}\frac{\partial^2}{\partial\alpha_3^2}\widetilde\Phi_{4;\pi}(\underline{\alpha}))\Big ) \, .
\end{eqnarray}
\end{widetext}
In the above expressions, $f_n(x) \equiv 1 - e^{-x} \sum_{k=0}^n {x^k \over k!}$; the parameters $\omega$ and $\omega^\prime$ are transformed to be $T_1$ and $T_2$, respectively; we choose $T_1 = T_2 = 2T$ so that $u_0 = {T_1 \over T_1 + T_2} = {1\over2}$; we choose $\omega_c = 1.62$~GeV to be the average threshold value of the $\Xi_b^-({3/2}^-)$ and $\Xi_b^{*0}(3/2^+)$ mass sum rules; we choose $0.268~\rm{GeV}<T<0.322~\rm{GeV}$ to be the Borel window of the $\Xi_b^-({3/2}^-)$ mass sum rule. The light-cone distribution amplitudes contained in the above sum rule expressions can be found in Refs.~\cite{Ball:1998je,Ball:2006wn,Ball:2004rg,Ball:1998kk,Ball:1998sk,Ball:1998ff,Ball:2007rt,Ball:2007zt}.

\begin{figure}[H]
\centering
\scalebox{0.6}{\includegraphics{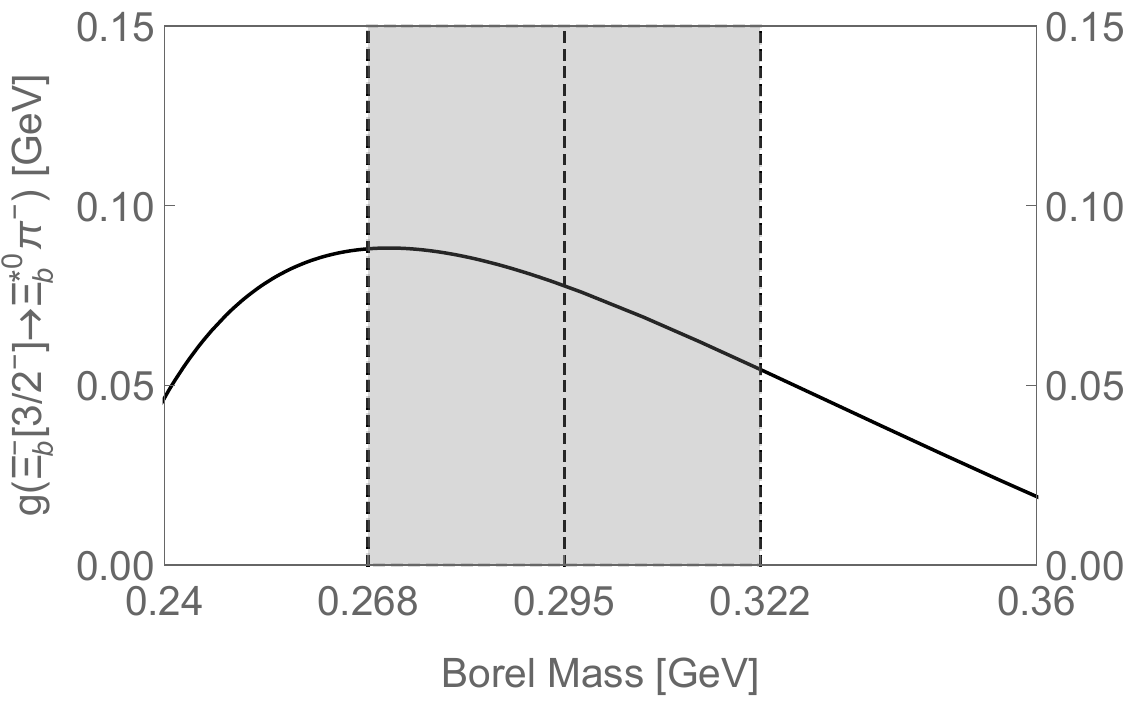}}
\caption{The coupling constant $g^S_{\Xi_b^0[{3\over2}^-] \rightarrow \Xi_b^{*+}\pi^-}$ as a function of the Borel mass $T$.}
\label{fig:g}
\end{figure}

We extract the coupling constant from Eq.~(\ref{eq:g}) to be:
\begin{eqnarray}
\nonumber g^S_{\Xi_b^-[{3\over2}^-] \rightarrow \Xi_b^{*0}\pi^-} &=& 0.08~{^{+0.01}_{-0.03}}~{^{+0.03}_{-0.03}}~{^{+003}_{-0.03}}~{^{+0.44}_{-0.08}}~{\rm GeV}^{-2}
\\ &=& 0.08{^{+0.44}_{-0.08}}~{\rm GeV}^{-2} \, ,
\end{eqnarray}
where the uncertainties are due to the Borel mass, parameters of $\Xi_b^{*0}(1/2^+)$, parameters of $\Xi_b^-({3/2}^-)$, and various QCD parameters given in Eqs.~(\ref{eq:condensates}), respectively. We depict $g^S_{\Xi_b^-[{3\over2}^-] \rightarrow \Xi_b^{*0}\pi^-}$ in Fig.~\ref{fig:g} as a function of the Borel mass $T$. Its dependence on $T$ is weak inside the Borel window $0.268<T<0.322$~GeV.

For the three-body decay process $\Xi_b(3/2^-)\to \Xi_b^*+ \pi\to \Xi_b+ \pi +\pi$ , we use the formula,
\begin{eqnarray}
&&\Gamma\left(0\to 4+3\to 3+1+2\right)
\\ \nonumber &\equiv& \left(\Xi_b^-(3/2^-)\to\Xi_b^*+\pi\to\Xi_b+\pi+\pi\right)
\\ \nonumber &=& {1\over(2\pi)^3}{1\over 32 m_0^3}\times g_{0\to 3+4}^2\times g_{4\to2+1}^2\times\int dm_{12} dm_{23}
\\ \nonumber &\times& {9\over2}\times {1\over4}\times{\rm Tr}\Big[(p\!\!\!\slash_1+m_1)(g_{\beta_2\alpha_2}-{1\over3}\gamma_{\beta_2}\gamma_{\alpha_2}
\\ \nonumber &-&{p_{4,\beta_2}\gamma_{\alpha_2}-p_{4,\alpha_2}\gamma_{\beta_2}\over 3m_4}-{2p_{4,\beta_2}p_{4,\alpha_2}\over 3m_4})(p\!\!\!\slash_4+m_4)
\\ \nonumber &\times&(g_{\alpha_1\alpha_2}-{1\over3}\gamma_{\alpha_2}\gamma_{\alpha_1}
- {p_{0,\alpha_2}\gamma_{\alpha_1}-p_{0,\alpha_1}\gamma_{\alpha_2}\over 3 m_0}-{2p_{0,\alpha_2}p_{0,\alpha_1}\over 3 m_0})
\\ \nonumber &\times&(p\!\!\!\slash_0+m_0)(g_{\alpha_1\beta_1}-{1\over3}\gamma_{\alpha_1}\gamma_{\beta_1}-{p_{4,\alpha_1}\gamma_{\beta_1}-p_{4,\beta_1}\gamma_{\alpha_1}\over 3m_4}
\\ \nonumber &-&{2p_{4,\alpha_1}p_{4,\beta_1}\over 3m_4})
(p\!\!\!\slash_4+m_4)\Big]\times {p_{2,\beta_1}p_{2,\beta_2}\over |p_4^2-m_4^2+i m_4\Gamma_4|^2}\, ,
\end{eqnarray}
to calculate its partial decay width to be
\begin{eqnarray}
\Gamma^S_{\Xi_b^-[{3\over2}^-] \rightarrow \Xi_b^{*0}\pi^-\to\Xi_b^0\pi^0\pi^-}&=& 0.07{^{+2.94}_{-0.07}}{\rm~MeV} \, .
\end{eqnarray}

Similarly, we study the other decay channels for the bottom baryons belonging to the $[\mathbf{\bar 3}_F, 1, 1, \rho]$ and $[\mathbf{\bar 3}_F, 1, 0, \lambda]$ doublets. Their partial decay widths are evaluated and summarized in Table~\ref{tab:decayb311rho}. Our results suggest that the $\Xi_b(6100)$ can be well explained as the $P$-wave bottom baryon of $J^P=3/2^-$ belonging to the $[\mathbf{\bar 3}_F,1,1,\rho]$ doublet, while the $[\mathbf{\bar 3}_F,1,0,\lambda]$ doublet is disfavored.

\begin{table*}[hbt]
\begin{center}
\renewcommand{\arraystretch}{1.5}
\caption{Decay properties of the $P$-wave bottom baryons belonging to the $[\mathbf{\bar 3}_F, 1, 1, \rho]$ and $[\mathbf{\bar 3}_F,1,0,\lambda]$ doublets, calculated using the method of light-cone sum rules within the framework of heavy quark effective theory. Possible experimental candidates are given in the last column.}
\setlength{\tabcolsep}{0.1mm}{
\begin{tabular}{c| c | c | c | c | c | c | c | c}
\hline\hline
\multirow{2}{*}{Multiplets}&Baryon & ~~~~Mass~~~~ & Difference & \multirow{2}{*}{~~~~~~~~~Decay channels~~~~~~~~~}  & ~$S$-wave width~  & ~$D$-wave width~ & ~Total width~ & \multirow{2}{*}{~Candidate~}
\\ & ($j^P$) & ({GeV})& ({MeV}) & & ({MeV}) & ({MeV}) & ({MeV})
\\ \hline\hline
\multirow{12}{*}{$[\mathbf{\bar 3}_F,1,1,\rho]$}&\multirow{3}{*}{$\Lambda_b({1\over2}^-)$}&\multirow{3}{*}{$5.92^{+0.13}_{-0.10}$}&\multirow{6}{*}{$10^{+4}_{-4}$}&$\Lambda_b({1\over2}^-)\to \Sigma_b\pi\to\Lambda_b\pi\pi$&$4\times10^{-3}$&--&\multirow{3}{*}{$4\times10^{-3}$}&\multirow{3}{*}{$\Lambda_b(5912)$}
\\ \cline{5-7}
&&&&$\Lambda_b({1\over2}^-)\to\Sigma_b^*\pi\to\Lambda_b\pi\pi$&--&$2\times10^{-8}$&&
\\ \cline{5-7}
&&&&$\Lambda_b({1\over2}^-)\to \Lambda_b\rho\to\Lambda_b\pi\pi$&\multicolumn{2}{c|}{$1\times10^{-4}$}&&
\\ \cline{2-3} \cline{5-9}
&\multirow{3}{*}{$\Lambda_b({3\over2}^-)$}&\multirow{3}{*}{$5.93^{+0.13}_{-0.10}$}&&$\Lambda_b({3\over2}^-)\to\Sigma_b^*\pi\to\Lambda_b\pi\pi$&$5\times10^{-5}$&$3\times10^{-8}$&\multirow{3}{*}{$1\times10^{-4}$}&\multirow{3}{*}{$\Lambda_b(5920)$}
\\ \cline{5-7}
&&&&$\Lambda_b({3\over2}^-)\to\Sigma_b\pi\to\Lambda_b\pi\pi$&--&$5\times10^{-7}$&&
\\ \cline{5-7}
&&&&$\Lambda_b({3\over2}^-)\to \Lambda_b\rho\to\Lambda\pi \pi$ & \multicolumn{2}{c|}{$7\times10^{-5}$}&&
\\ \cline{2-9}
&\multirow{3}{*}{$\Xi_b({1\over2}^-)$}&\multirow{3}{*}{$6.08^{+0.13}_{-0.11}$}&\multirow{7}{*}{$9^{+3}_{-3}$}&$\Xi_c({1\over2}^-)\to \Xi_b^{\prime}\pi$&$3.6^{+29.4}_{-~3.6}$&--&\multirow{3}{*}{$4^{+29}_{-~4}$}&\multirow{4}{*}{--}
\\ \cline{5-7}
&&&&$\Xi_b({1\over2}^-)\to\Xi_b^{*}\pi\to\Xi_b\pi\pi$&--&$2\times10^{-6}$&
\\ \cline{5-7}
&&&&$\Xi_b({1\over2}^-)\to\Xi_b\rho\to\Xi_c\pi\pi$&\multicolumn{2}{c|}{$2\times10^{-4}$}&
\\ \cline{2-3} \cline{5-9}
&\multirow{3}{*}{$\Xi_b({3\over2}^-)$}&\multirow{3}{*}{$6.09^{+0.12}_{-0.11}$}&&$\Xi_b({3\over2}^-)\to\Xi_ b^{\prime}\pi$&--&$2\times10^{-3}$&\multirow{3}{*}{$0.1^{+3.0}_{-0.1}$}&\multirow{3}{*}{$\Xi_b(6100)$}
\\ \cline{5-7}
&&&&$\Xi_b({3\over2}^-)\to\Xi_b^{*}\pi\to\Xi_b\pi\pi$&$0.07^{+2.94}_{-0.07}$&$2\times10^{-5}$&
\\ \cline{5-7}
&&&&$\Xi_b({3\over2}^-)\to\Xi_b \rho\to\Xi_b \pi \pi$&\multicolumn{2}{c|}{$2\times10^{-4}$}&
\\ \hline\hline
\multirow{12}{*}{$[\mathbf{\bar 3}_F,1,0,\lambda]$}&\multirow{3}{*}{$\Lambda_b({1\over2}^-)$}&\multirow{3}{*}{$5.91^{+0.11}_{-0.11}$}&\multirow{6}{*}{$6^{+2}_{-2}$}&$\Lambda_b({1\over2}^-)\to \Sigma_b\pi\to\Lambda_b\pi\pi$&$0.92^{+~1.5}_{-0.66}$&--&\multirow{3}{*}{$0.9^{+1.5}_{-0.7}$}&\multirow{3}{*}{$\Lambda_b(5912)$}
\\ \cline{5-7}
&&&&$\Lambda_b({1\over2}^-)\to \Sigma_b^*\pi\to\Lambda_b\pi\pi$&--&$2\times10^{-7}$&&
\\ \cline{5-7}
&&&&$\Lambda_b({1\over2}^-)\to \Lambda_b\rho\to\Lambda_b\pi\pi$&\multicolumn{2}{c|}{$3\times10^{-4}$}&&
\\ \cline{2-3} \cline{5-9}
&\multirow{3}{*}{$\Lambda_b({3\over2}^-)$}&\multirow{3}{*}{$5.92^{+0.11}_{-0.11}$}&&$\Lambda_b({3\over2}^-)\to\Sigma_b^*\pi\to\Lambda_b\pi\pi$&$0.56^{+0.75}_{-0.37}$&$2\times10^{-7}$&\multirow{3}{*}{$0.6^{+0.8}_{-0.4}$}&\multirow{3}{*}{$\Lambda_b(5920)$}
\\ \cline{5-7}
&&&&$\Lambda_b({3\over2}^-)\to\Sigma_b\pi\to\Lambda_b\pi\pi$&--&$3\times10^{-6}$&&
\\ \cline{5-7}
&&&&$\Lambda_b({3\over2}^-)\to \Lambda_b\rho\to\Lambda\pi \pi$ & \multicolumn{2}{c|}{$7\times10^{-4}$} &&
\\ \cline{2-9}
&\multirow{3}{*}{$\Xi_b({1\over2}^-)$}&\multirow{3}{*}{$6.09^{+0.10}_{-0.10}$}&\multirow{6}{*}{$5^{+2}_{-2}$}& $\Xi_b({1\over2}^-)\to \Xi_b^{\prime}\pi$&$1200^{+1600}_{-~850}$&--&\multirow{3}{*}{$1200^{+1600}_{-~850}$}&\multirow{3}{*}{--}
\\ \cline{5-7}
&&&&$\Xi_b({1\over2}^-)\to\Xi_b^*\pi\to\Xi_b\pi\pi$&--&$1\times10^{-5}$&&
\\ \cline{5-7}
&&&&$\Xi_b({1\over2}^-)\to\Xi_b\rho\to\Xi_b\pi\pi$&\multicolumn{2}{c|}{$6\times10^{-4}$}&
\\ \cline{2-3} \cline{5-9}
&\multirow{3}{*}{$\Xi_b({3\over2}^-)$}&\multirow{3}{*}{$6.10^{+0.10}_{-0.10}$}&&$\Xi_b({3\over2}^-)\to\Xi_ b^{\prime}\pi$&--&$8\times10^{-3}$&\multirow{3}{*}{$240^{+290}_{-150}$}&\multirow{4}{*}{--}
\\ \cline{5-7}
&&&&$\Xi_b({3\over2}^-)\to\Xi_b^{*}\pi\to\Xi_b\pi\pi$&$240^{+290}_{-150}$&$8\times 10^{-5}$&
\\ \cline{5-7}
&&&&$\Xi_b({3\over2}^-)\to\Xi_b \rho\to\Xi_b \pi \pi$&\multicolumn{2}{c|}{$1\times10^{-3}$}&
\\ \hline\hline
\end{tabular}}
\label{tab:decayb311rho}
\end{center}
\end{table*}

%
\section{Summary and Discussions}\label{sec:summary}
%

In this paper we study the $\Xi_b(6100)$ as a possible $P$-wave bottom baryon of $J^P=3/2^-$ belonging to the $[\mathbf{\bar 3}_F,1,1,\rho]$ doublet. This doublet contains four $P$-wave bottom baryons: $\Lambda_b(1/2^-)$, $\Lambda_b(3/2^-)$, $\Xi_b(1/2^-)$, and $\Xi_b(3/2^-)$. We calculate their masses using the QCD sum rule method within the framework of heavy quark effective theory. We also study their $S$- and $D$-wave decays into the ground-state bottom baryons and light pseudoscalar/vector mesons through the light-cone sum rule method. The obtained results are summarized in Table~\ref{tab:decayb311rho}, where the masses and total widths are evaluated to be:
\begin{eqnarray}
\nonumber M_{\Lambda_b(1/2^-)}&=& 5.92^{+0.13}_{-0.10}~\mbox{MeV}\, ,
\\ \nonumber \Gamma_{\Lambda_b(1/2^-)}&\sim& 0 \, ,
\\ \nonumber M_{\Lambda_b(3/2^-)}&=& 5.93^{+0.13}_{-0.10}~\mbox{MeV}\, ,
\\ \nonumber \Gamma_{\Lambda_b(3/2^-)}&\sim& 0\, ,
\\ M_{\Lambda_b(3/2^-)}-M_{\Lambda_b(1/2^-)}&=&10\pm 4~\mbox{MeV}\, ,
\\ \nonumber M_{\Xi_b(1/2^-)}&=& 6.08^{+0.13}_{-0.11}~\mbox{MeV}\, ,
\\ \nonumber \Gamma_{\Xi_b(1/2^-)}&=& 4^{+29}_{-~4}~\mbox{MeV}\, ,
\\ \nonumber M_{\Xi_b(3/2^-)}&=& 6.09^{+0.12}_{-0.11}~\mbox{MeV}\, ,
\\ \nonumber \Gamma_{\Xi_b(3/2^-)}&=& 0.1^{+3.0}_{-0.1}~\mbox{MeV}\, ,
\\ \nonumber  M_{\Xi_b(3/2^-)}-M_{\Xi_b(1/2^-)}&=& 9\pm 3~\mbox{MeV}\, .
\end{eqnarray}
Our results suggest that the $\Xi_b(6100)$ can be well explained as the $P$-wave bottom baryon of $J^P=3/2^-$ belonging to the $[\mathbf{\bar 3}_F,1,1,\rho]$ doublet. The $\Lambda_b(5912)$ and $\Lambda_b(5920)$ can be explained as its partner states with $J^P=1/2^-$ and $3/2^-$, respectively. Besides, there is a $P$-wave bottom baryon still missing, the $\Xi_b(1/2^-)$, whose mass is $\Delta M=9\pm3$~MeV smaller than the $\Xi_b(6100)$. We propose to search for it in the $\Xi_c({1/2}^-)\to \Xi_b^{\prime}\pi$ decay channel.

Besides the $[\mathbf{\bar 3}_F, 1, 1, \rho]$ doublet, the $\Xi_b(6100)$ may also belong to the $[\mathbf{\bar 3}_F, 1, 0, \lambda]$ doublet. In the present study we have also investigated the latter doublet, and the obtained results are also summarized in Table~\ref{tab:decayb311rho}. These results can also be used to explain the $\Lambda_b(5912)$ and $\Lambda_b(5920)$, so our QCD sum rule results can not distinguish whether they belong to the $[\mathbf{\bar 3}_F, 1, 1, \rho]$ doublet or the $[\mathbf{\bar 3}_F,1,0,\lambda]$ doublet. This is partly because that their decays into the $\Sigma_b^{(*)} \pi$ channels are kinematically forbidden so that their widths are limited. However, our results for the the $[\mathbf{\bar 3}_F, 1, 0, \lambda]$ doublet can not be used to easily explain the $\Xi_b(6100)$. We would like to note that the above results are just possible explanations, and there exist many other possibilities. Especially, our QCD sum rule results for the $\Xi_b(6100)$ seem to be not consistent with the quark model expectation~\cite{Yoshida:2015tia,Nagahiro:2016nsx}, so further experimental and theoretical studies are crucially demanded to fully understand them. There exists a relevant question on how to explain the five excited $\Omega_c$ baryons observed by LHCb~\cite{LHCb:2017uwr}, given that at most four of them can be explained as the $P$-wave excitations of the $\lambda$-mode~\cite{Yang:2021lce,Chen:2017gnu}. There are two possible assignments for the rest of them: either the radial $2S$-wave excitation or the orbital $1P$-wave excitation of the $\rho$-mode. The experimental measurements on the quantum numbers of these excited $\Omega_c$ baryons can be important and helpful to understand the $\rho$-mode excitations.

For completeness, we have also investigated the $P$-wave charmed baryons belonging to the $[\mathbf{\bar 3}_F, 1, 1, \rho]$ and $[\mathbf{\bar 3}_F, 1, 0, \lambda]$ doublets. The obtained results are summarized in Appendix~\ref{sec:charmedsumrule}, suggesting that the $\Lambda_c(2595)$, $\Xi_c(2790)$, $\Lambda_c(2625)$, and $\Xi_c(2815)$, as the charmed partner states of the $\Lambda_b(5912)$, $\Lambda_b(5920)$, $\Xi_b(1/2^-)$, and $\Xi_b(6100)$, can be well explained as the $P$-wave charmed baryons belonging to the $[\mathbf{\bar 3}_F,1,1,\rho]$ doublet.

%
\section*{Acknowledgments}

This project is supported by
the National Natural Science Foundation of China under Grant No.~12075019,
the Jiangsu Provincial Double-Innovation Program under Grant No.~JSSCRC2021488,
and
the Fundamental Research Funds for the Central Universities.

\appendix

\section{$P$-wave charmed baryons from $[\mathbf{\bar 3}_F, 1, 1, \rho]$ and $[\mathbf{\bar 3}_F, 1, 0, \lambda]$}
\label{sec:charmedsumrule}

In this appendix we study the $P$-wave charmed baryons belonging to the $[\mathbf{\bar 3}_F, 1, 1, \rho]$ and $[\mathbf{\bar 3}_F, 1, 0, \lambda]$ doublets. We apply the QCD sum rule method to study the mass spectrum, and the obtained results are summarized in Table~\ref{tab:pwavecparameter}. We apply the light-cone sum rule method to study the decay properties, and the obtained results are summarized in Table~\ref{tab:decayc311rho}. These results suggest that the $\Lambda_c(2595)$, $\Xi_c(2790)$, $\Lambda_c(2625)$, and $\Xi_c(2815)$, as the charmed partner states of the $\Lambda_b(5912)$, $\Lambda_b(5920)$, $\Xi_b(1/2^-)$, and $\Xi_b(6100)$, can be well explained as the $P$-wave charmed baryons belonging to the $[\mathbf{\bar 3}_F,1,1,\rho]$ doublet.

\begin{table*}[hbtp]
\begin{center}
\renewcommand{\arraystretch}{1.5}
\caption{Parameters of the $P$-wave charmed baryons belonging to the $[\mathbf{\bar 3}_F, 1, 1, \rho]$ and $[\mathbf{\bar 3}_F, 1, 0, \lambda]$ doublets, calculated using the method of QCD sum rules within the framework of heavy quark effective theory.}
\begin{tabular}{c| c | c | c | c | c c | c | c}
\hline\hline
\multirow{2}{*}{Multiplets}&\multirow{2}{*}{~~B~~} & $\omega_c$ & ~~~Working region~~~ & ~~~~~~~$\overline{\Lambda}$~~~~~~~ & ~~~Baryon~~~ & ~~~~Mass~~~~~ & ~Difference~ & Decay constant
\\                                         &      & (GeV)      &(GeV)                & (GeV)                              & ($j^P$)       & (GeV)      & (MeV)        & (GeV$^{4}$)
\\ \hline\hline
\multirow{4}{*}{$[\mathbf{\bar 3}_F, 1, 1, \rho]$}& \multirow{2}{*}{$\Lambda_c$} & \multirow{2}{*}{$1.53$} & \multirow{2}{*}{$0.27< T < 0.29$} & \multirow{2}{*}{$1.16 \pm 0.09$} & $\Lambda_b(1/2^-)$ & $2.59 ^{+0.14}_{-0.13}$ & \multirow{2}{*}{$51 ^{+20}_{-18}$} & $0.051 \pm 0.012~(\Lambda^+_c(1/2^-))$
\\ \cline{6-7}\cline{9-9}
& & & & & $\Lambda_c(3/2^-)$ & $2.64 ^{+0.13}_{-0.12}$ & &$0.024 \pm 0.006~(\Lambda^+_c(3/2^-))$
\\ \cline{2-9}
&\multirow{2}{*}{$\Xi_c$} & \multirow{2}{*}{$1.78$} & \multirow{2}{*}{$0.27< T < 0.32$} & \multirow{2}{*}{$1.33 \pm 0.10$} & $\Xi_c(1/2^-)$ & $2.78 ^{+0.17}_{-0.16}$ & \multirow{2}{*}{$42 ^{+17}_{-15}$} & $0.076 \pm 0.019~(\Xi_c^0(1/2^-))$
\\ \cline{6-7}\cline{9-9}
& & & & & $\Xi_c(3/2^-)$ & $2.82 ^{+0.15}_{-0.13}$ & &$0.036 \pm 0.009~(\Xi_c^0(3/2^-))$
\\ \hline \hline
\multirow{4}{*}{$[\mathbf{\bar 3}_F, 1, 0, \lambda]$}& \multirow{2}{*}{$\Lambda_c$} & \multirow{2}{*}{$1.43$}&\multirow{2}{*}{$T=0.30$} & \multirow{2}{*}{$0.95 \pm 0.07$} & $\Lambda_b(1/2^-)$ & $2.65^{+0.09}_{-0.07}$ & \multirow{2}{*}{$33^{+12}_{-12}$} & $0.019 \pm 0.004~(\Lambda^+_c(1/2^-))$
\\ \cline{6-7}\cline{9-9}
& & & & & $\Lambda_c(3/2^-)$ & $2.68^{+0.09}_{-0.07}$ & &$0.009 \pm 0.002~(\Lambda^+_c(3/2^-))$
\\ \cline{2-9}
&\multirow{2}{*}{$\Xi_c$} & \multirow{2}{*}{$1.68$} & \multirow{2}{*}{$T=0.33$} & \multirow{2}{*}{$1.15 \pm 0.09$} & $\Xi_c(1/2^-)$ & $2.91^{+ 0.12}_{-0.11}$ & \multirow{2}{*}{$27 ^{+10}_{-~9}$} & $0.032 \pm 0.006~(\Xi_c^0(1/2^-))$
\\ \cline{6-7}\cline{9-9}
& & & & & $\Xi_c(3/2^-)$ & $2.94^{+ 0.12}_{-0.11}$ & &$0.015 \pm 0.003~(\Xi_c^0(3/2^-))$
\\ \hline \hline
\end{tabular}
\label{tab:pwavecparameter}
\end{center}
\end{table*}

\begin{table*}[hbt]
\begin{center}
\renewcommand{\arraystretch}{1.5}
\caption{Decay properties of the $P$-wave charmed baryons belonging to the $[\mathbf{\bar 3}_F, 1, 1, \rho]$ and $[\mathbf{\bar 3}_F, 1, 0, \lambda]$ doublets, calculated using the method of light-cone sum rules within the framework of heavy quark effective theory. Possible experimental candidates are given in the last column.}
\setlength{\tabcolsep}{0.1mm}{
\begin{tabular}{c| c | c | c | c | c | c | c | c}
\hline\hline
\multirow{2}{*}{Multiplets}&Baryon & ~~~~Mass~~~~ & Difference & \multirow{2}{*}{~~~~~~~~~Decay channels~~~~~~~~~}  & ~$S$-wave width~  & ~$D$-wave width~ & ~Total width~ & \multirow{2}{*}{~Candidate~}
\\ & ($j^P$) & ({GeV})& ({MeV}) & & ({MeV}) & ({MeV}) & ({MeV})
\\ \hline\hline
\multirow{12}{*}{$[\mathbf{\bar 3}_F,1,1,\rho]$}&\multirow{3}{*}{$\Lambda_c({1\over2}^-)$}& \multirow{3}{*}{$2.59^{+0.14}_{-0.13}$} &\multirow{6}{*}{$51^{+20}_{-18}$}&$\Lambda_c({1\over2}^-)\to \Sigma_c\pi\to\Lambda_c\pi\pi$&$5.4^{+40.2}_{-~4.7}$&--&\multirow{3}{*}{$5^{+40}_{-~5}$}&\multirow{3}{*}{$\Lambda_c(2595)$}
\\ \cline{5-7}
&&&&$\Lambda_c({1\over2}^-)\to\Sigma_c^*\pi\to\Lambda_c\pi\pi$&--&$6\times10^{-8}$&&
\\ \cline{5-7}
&&&&$\Lambda_c({1\over2}^-)\to \Lambda_c\rho\to \Lambda_c\pi\pi$& \multicolumn{2}{c|}{$8\times10^{-4}$ }&
\\ \cline{2-3} \cline{5-9}
&\multirow{3}{*}{$\Lambda_c({3\over2}^-)$}&\multirow{3}{*}{$2.64^{+0.13}_{-0.12}$}&&$\Lambda_c({3\over2}^-)\to \Sigma_c \pi$ & -- & $4\times10^{-3}$&\multirow{3}{*}{$0.1^{+0.5}_{-0.1}$}&\multirow{3}{*}{$\Lambda_c(2625)$}
\\ \cline{5-7}
&&&&$\Lambda_c({3\over2}^-)\to\Sigma_c^*\pi\to\Lambda_c\pi\pi$&$0.06^{+0.53}_{-0.06}$&$9\times10^{-7}$&&
\\ \cline{5-7}
&&&&$\Lambda_c({3\over2}^-)\to \Lambda_c\rho\to \Lambda_c\pi\pi$&\multicolumn{2}{c|}{$0.01^{+0.03}_{-0.01}$}&
\\ \cline{2-9}
&\multirow{3}{*}{$\Xi_c({1\over2}^-)$}&\multirow{3}{*}{$2.78^{+0.17}_{-0.16}$}&\multirow{8}{*}{$42^{+17}_{-15}$}&$\Xi_c({1\over2}^-)\to \Xi_c^{\prime}\pi$&$9.0^{+59.8}_{-~9.0}$&--&\multirow{3}{*}{$9^{+60}_{-~9}$}&\multirow{4}{*}{$\Xi_c(2790)$}
\\ \cline{5-7}
&&&&$\Xi_c({1\over2}^-)\to\Xi_c^*\pi\to\Xi_c\pi\pi$&--&$1\times10^{-5}$&&
\\ \cline{5-7}
&&&&$\Xi_c({1\over2}^-)\to\Xi_c\rho\to\Xi_c\pi\pi$&\multicolumn{2}{c|}{$2\times10^{-3}$}&&
\\ \cline{2-3} \cline{5-9}
&\multirow{3}{*}{$\Xi_c({3\over2}^-)$}&\multirow{3}{*}{$2.82^{+0.15}_{-0.13}$}&&$\Xi_c({3\over2}^-)\to\Xi_ c^{*}\pi\to\Xi_c\pi\pi$&$4.3^{+49.6}_{-~4.3}$&$1\times10^{-4}$&\multirow{3}{*}{$4^{+50}_{-~4}$}&\multirow{3}{*}{$\Xi_c(2815)$}
\\ \cline{5-7}
&&&&$\Xi_c({3\over2}^-)\to\Xi_c^{\prime}\pi$&--&$0.03$&&
\\ \cline{5-7}
&&&&$\Xi_c({3\over2}^-)\to \Xi_c\rho\to\Xi_c\pi\pi$&\multicolumn{2}{c|}{$0.01$}&&
\\ \hline \hline
\multirow{12}{*}{$[\mathbf{\bar 3}_F,1,0,\lambda]$}&\multirow{3}{*}{$\Lambda_c({1\over2}^-)$}& \multirow{3}{*}{$2.65^{+0.09}_{-0.07}$} &\multirow{6}{*}{$33^{+12}_{-12}$}&$\Lambda_c({1\over2}^-)\to \Sigma_c\pi$&$>560$&--&\multirow{3}{*}{$>560$}&\multirow{3}{*}{--}
\\ \cline{5-7}
&&&&$\Lambda_c({1\over2}^-)\to\Sigma_c^*\pi\to\Lambda_c\pi\pi$&--&$2\times10^{-8}$&&
\\ \cline{5-7}
&&&&$\Lambda_c({1\over2}^-)\to \Lambda_c\rho\to \Lambda_c\pi\pi$& \multicolumn{2}{c|}{$2\times 10^{-3}$ }&
\\ \cline{2-3} \cline{5-9}
&\multirow{3}{*}{$\Lambda_c({3\over2}^-)$}&\multirow{3}{*}{$2.68^{+0.09}_{-0.07}$}&&$\Lambda_c({3\over2}^-)\to \Sigma_c \pi$ & -- & $0.05$&\multirow{3}{*}{$35^{+49}_{-26}$}&\multirow{3}{*}{--}
\\ \cline{5-7}
&&&&$\Lambda_c({3\over2}^-)\to\Sigma_c^*\pi\to\Lambda_c\pi\pi$&$35^{+49}_{-26}$&$2\times10^{-5}$&&
\\ \cline{5-7}
&&&&$\Lambda_c({3\over2}^-)\to \Lambda_c\rho\to \Lambda_c\pi\pi$&\multicolumn{2}{c|}{$0.01$}&
\\ \cline{2-9}
&\multirow{3}{*}{$\Xi_c({1\over2}^-)$}&\multirow{3}{*}{$2.91^{+0.12}_{-0.11}$}&\multirow{7}{*}{$27^{+10}_{-~9}$}&$\Xi_c({1\over2}^-)\to \Xi_c^{\prime}\pi$&$>1500$&--&\multirow{3}{*}{$>1500$}&\multirow{3}{*}{--}
\\ \cline{5-7}
&&&&$\Xi_c({1\over2}^-)\to\Xi_c^*\pi\to\Xi_c\pi\pi$&--&$1\times10^{-4}$&&
\\ \cline{5-7}
&&&&$\Xi_c({1\over2}^-)\to\Xi_c\rho\to\Xi_c\pi\pi$&\multicolumn{2}{c|}{$5\times10^{-3}$}&&
\\ \cline{2-3} \cline{5-9}
&\multirow{3}{*}{$\Xi_c({3\over2}^-)$}&\multirow{3}{*}{$2.94^{+0.12}_{-0.11}$}&&$\Xi_c({3\over2}^-)\to\Xi_ c^{*}\pi\to\Xi_b\pi\pi$&$>230$&$9\times10^{-4}$&\multirow{3}{*}{$>230$}&\multirow{3}{*}{--}
\\ \cline{5-7}
&&&&$\Xi_c({3\over2}^-)\to\Xi_c^{\prime}\pi$&--&$0.16^{+0.23}_{-0.11}$&&
\\ \cline{5-7}
&&&&$\Xi_c({3\over2}^-)\to \Xi_c\rho\to\Xi_c\pi\pi$&\multicolumn{2}{c|}{$0.02$}&&
\\ \hline\hline
\end{tabular}}
\label{tab:decayc311rho}
\end{center}
\end{table*}

%

%


\begin{thebibliography}{96}

\bibitem{Korner:1994nh}
J.~G.~Korner, M.~Kramer and D.~Pirjol,
{\it Heavy baryons},
Prog. Part. Nucl. Phys. \textbf{33} (1994), 787-868.

\bibitem{Manohar:2000dt}
A.~V.~Manohar and M.~B.~Wise,
{\it Heavy quark physics},
Camb. Monogr. Part. Phys. Nucl. Phys. Cosmol. \textbf{10} (2000), 1-191.

\bibitem{Bianco:2003vb}
S.~Bianco, F.~L.~Fabbri, D.~Benson and I.~Bigi,
{\it A Cicerone for the physics of charm},
Riv. Nuovo Cim. \textbf{26} (2003) no.7-8, 1-200.

\bibitem{Klempt:2009pi}
E.~Klempt and J.~M.~Richard,
{\it Baryon spectroscopy},
Rev. Mod. Phys. \textbf{82} (2010), 1095-1153.

\bibitem{Kalman:1983an}
C.~S.~Kalman and D.~Pfeffer,
{\it Ground State and $P$ Wave $B$ Flavored Baryons in a Consistent Quark Model With Hyperfine Interactions},
Phys. Rev. D \textbf{28} (1983), 2324.

\bibitem{Isgur:1978wd}
N.~Isgur and G.~Karl,
{\it Positive Parity Excited Baryons in a Quark Model with Hyperfine Interactions},
Phys. Rev. D \textbf{19} (1979), 2653,
[erratum: Phys. Rev. D \textbf{23} (1981), 817].

\bibitem{Copley:1979wj}
L.~A.~Copley, N.~Isgur and G.~Karl,
{\it Charmed Baryons in a Quark Model with Hyperfine Interactions},
Phys. Rev. D \textbf{20} (1979), 768,
[erratum: Phys. Rev. D \textbf{23} (1981), 817].

\bibitem{Karliner:2008sv}
M.~Karliner, B.~Keren-Zur, H.~J.~Lipkin and J.~L.~Rosner,
{\it The Quark Model and $b$ Baryons},
Annals Phys. \textbf{324} (2009), 2-15.

\bibitem{Chen:2016spr}
H.~X.~Chen, W.~Chen, X.~Liu, Y.~R.~Liu and S.~L.~Zhu,
{\it A review of the open charm and open bottom systems},
Rept. Prog. Phys. \textbf{80} (2017) no.7, 076201.

\bibitem{Chen:2022asf}
H.~X.~Chen, W.~Chen, X.~Liu, Y.~R.~Liu and S.~L.~Zhu,
{\it An updated review of the new hadron states},
[arXiv:2204.02649 [hep-ph]].

\bibitem{Knapp:1976qw}
B.~Knapp, W.~Y.~Lee, P.~Leung, S.~D.~Smith, A.~Wijangco, J.~Knauer, D.~Yount, J.~Bronstein, R.~Coleman and G.~Gladding, \textit{et al.}
{\it Observation of a Narrow anti-Baryon State at $2.26~ GeV/c^2$},
Phys. Rev. Lett. \textbf{37} (1976), 882.

\bibitem{pdg}
M.~Tanabashi \textit{et al.} [Particle Data Group],
{\it Review of Particle Physics},
Phys. Rev. D \textbf{98}, no.3, 030001 (2018).

\bibitem{ARGUS:1993vtm}
H.~Albrecht \textit{et al.} [ARGUS],
{\it Observation of a new charmed baryon},
Phys. Lett. B \textbf{317} (1993), 227-232.

\bibitem{E687:1993bax}
P.~L.~Frabetti \textit{et al.} [E687],
{\it An Observation of an excited state of the $\Lambda_c^+$ baryon},
Phys. Rev. Lett. \textbf{72} (1994), 961-964.

\bibitem{CLEO:1994oxm}
K.~W.~Edwards \textit{et al.} [CLEO],
{\it Observation of excited baryon states decaying to $\Lambda_c^+ \pi^+ \pi^-$},
Phys. Rev. Lett. \textbf{74} (1995), 3331-3335.

\bibitem{CLEO:1999msf}
J.~P.~Alexander \textit{et al.} [CLEO],
{\it Evidence of new states decaying into $\Xi_c^* \pi$},
Phys. Rev. Lett. \textbf{83} (1999), 3390-3393.

\bibitem{LHCb:2012kxf}
R.~Aaij \textit{et al.} [LHCb],
{\it Observation of excited $\Lambda_b^0$ baryons},
Phys. Rev. Lett. \textbf{109} (2012), 172003.

\bibitem{CMS:2021rvl}
A.~M.~Sirunyan \textit{et al.} [CMS],
{\it Observation of a New Excited Beauty Strange Baryon Decaying to $\Xi^-_\mathrm{b} \pi^+ \pi^-$},
Phys. Rev. Lett. \textbf{126} (2021) no.25, 252003.

\bibitem{Roberts:2007ni}
W.~Roberts and M.~Pervin,
{\it Heavy baryons in a quark model},
Int. J. Mod. Phys. A \textbf{23} (2008), 2817-2860.

\bibitem{Bijker:2020tns}
R.~Bijker, H.~Garc\'\i{}a-Tecocoatzi, A.~Giachino, E.~Ortiz-Pacheco and E.~Santopinto,
{\it Masses and decay widths of $\Xi_{c/b}$ and $\Xi^\prime_{c/b}$ baryons},
Phys. Rev. D \textbf{105} (2022) no.7, 074029.

\bibitem{Chen:2018vuc}
B.~Chen and X.~Liu,
{\it Assigning the newly reported $\Sigma_b(6097)$ as a $P$-wave excited state and predicting its partners},
Phys. Rev. D \textbf{98} (2018) no.7, 074032.

\bibitem{Chen:2018orb}
B.~Chen, K.~W.~Wei, X.~Liu and A.~Zhang,
{\it Role of newly discovered $\Xi_b(6227)^-$ for constructing excited bottom baryon family},
Phys. Rev. D \textbf{98} (2018) no.3, 031502.

\bibitem{Wang:2018fjm}
K.~L.~Wang, Q.~F.~L\"u and X.~H.~Zhong,
{\it Interpretation of the newly observed $\Sigma_b(6097)^{\pm}$ and $\Xi_b(6227)^-$ states as the $P$-wave bottom baryons},
Phys. Rev. D \textbf{99} (2019) no.1, 014011.

\bibitem{Xiao:2020gjo}
L.~Y.~Xiao and X.~H.~Zhong,
{\it Toward establishing the low-lying $P$-wave $\Sigma_b$ states},
Phys. Rev. D \textbf{102} (2020) no.1, 014009.

\bibitem{Ebert:2011kk}
D.~Ebert, R.~N.~Faustov and V.~O.~Galkin,
{\it Spectroscopy and Regge trajectories of heavy baryons in the relativistic quark-diquark picture},
Phys. Rev. D \textbf{84} (2011), 014025.

\bibitem{Chen:2016iyi}
B.~Chen, K.~W.~Wei, X.~Liu and T.~Matsuki,
{\it Low-lying charmed and charmed-strange baryon states},
Eur. Phys. J. C \textbf{77} (2017) no.3, 154.

\bibitem{Yang:2018lzg}
P.~Yang, J.~J.~Guo and A.~Zhang,
{\it Identification of the newly observed $\Sigma_b(6097)^\pm$ baryons from their strong decays},
Phys. Rev. D \textbf{99} (2019) no.3, 034018.

\bibitem{Lu:2020ivo}
Q.~F.~L\"u,
{\it Canonical interpretations of the newly observed $\Xi _c(2923)^0$, $\Xi _c(2939)^0$, and $\Xi _c(2965)^0$ resonances},
Eur. Phys. J. C \textbf{80} (2020) no.10, 921.

\bibitem{Wang:2020gkn}
K.~L.~Wang, L.~Y.~Xiao and X.~H.~Zhong,
{\it Understanding the newly observed $\Xi_c^0$ states through their decays},
Phys. Rev. D \textbf{102} (2020) no.3, 034029.

\bibitem{Wang:2019uaj}
K.~L.~Wang, Q.~F.~L\"u and X.~H.~Zhong,
{\it Interpretation of the newly observed $\Lambda_b(6146)^{0}$ and $\Lambda_b(6152)^0$ states in a chiral quark model},
Phys. Rev. D \textbf{100} (2019) no.11, 114035.

\bibitem{Lu:2014ina}
J.~X.~Lu, Y.~Zhou, H.~X.~Chen, J.~J.~Xie and L.~S.~Geng,
{\it Dynamically generated $J^P=1/2^-(3/2^-)$ singly charmed and bottom heavy baryons},
Phys. Rev. D \textbf{92} (2015) no.1, 014036.

\bibitem{Cheng:2015naa}
H.~Y.~Cheng and C.~K.~Chua,
{\it Strong Decays of Charmed Baryons in Heavy Hadron Chiral Perturbation Theory: An Update},
Phys. Rev. D \textbf{92} (2015) no.7, 074014.
\bibitem{Zeng:2020och}
C.~H.~Zeng, J.~X.~Lu, E.~Wang, J.~J.~Xie and L.~S.~Geng,
{\it Theoretical study of the $\Omega(2012)$ state in the $\Omega_c^0 \to \pi^+ \Omega(2012)^- \to \pi^+ (\bar{K}\Xi)^-$ and $\pi^+ (\bar{K}\Xi\pi)^-$ decays},
Phys. Rev. D \textbf{102} (2020) no.7, 076009.

\bibitem{Huang:2018bed}
Y.~Huang, C.~j.~Xiao, L.~S.~Geng and J.~He,
{\it Strong decays of the $\Xi_b(6227)$ as a $\Sigma_b\bar{K}$ molecule},
Phys. Rev. D \textbf{99} (2019) no.1, 014008.

\bibitem{Liang:2017ejq}
W.~H.~Liang, J.~M.~Dias, V.~R.~Debastiani and E.~Oset,
{\it Molecular $\Omega_b$ states},
Nucl. Phys. B \textbf{930} (2018), 524-532.

\bibitem{Chen:2017xat}
R.~Chen, A.~Hosaka and X.~Liu,
{\it Searching for possible $\Omega_c$-like molecular states from meson-baryon interaction},
Phys. Rev. D \textbf{97} (2018) no.3, 036016.

\bibitem{Liang:2014eba}
W.~H.~Liang, C.~W.~Xiao and E.~Oset,
{\it Baryon states with open beauty in the extended local hidden gauge approach},
Phys. Rev. D \textbf{89} (2014) no.5, 054023.

\bibitem{An:2017lwg}
C.~S.~An and H.~Chen,
{\it Observed $\Omega_{c}^{0}$ resonances as pentaquark states},
Phys. Rev. D \textbf{96} (2017) no.3, 034012.

\bibitem{Debastiani:2017ewu}
V.~R.~Debastiani, J.~M.~Dias, W.~H.~Liang and E.~Oset,
{\it Molecular $\Omega_c$ states generated from coupled meson-baryon channels},
Phys. Rev. D \textbf{97} (2018) no.9, 094035.

\bibitem{Guo:2008he}
X.~H.~Guo, K.~W.~Wei and X.~H.~Wu,
{\it Some mass relations for mesons and baryons in Regge phenomenology},
Phys. Rev. D \textbf{78} (2008), 056005.

\bibitem{Chen:2007xf}
C.~Chen, X.~L.~Chen, X.~Liu, W.~Z.~Deng and S.~L.~Zhu,
{\it Strong decays of charmed baryons},
Phys. Rev. D \textbf{75} (2007), 094017.

\bibitem{Ye:2017yvl}
D.~D.~Ye, Z.~Zhao and A.~Zhang,
{\it Study of $P$-wave excitations of observed charmed strange baryons},
Phys. Rev. D \textbf{96} (2017) no.11, 114009.


\bibitem{Chen:2014nyo}
B.~Chen, K.~W.~Wei and A.~Zhang,
{\it Assignments of $\Lambda_Q$ and $\Xi_Q$ baryons in the heavy quark-light diquark picture},
Eur. Phys. J. A \textbf{51} (2015), 82.
\bibitem{Aliev:2018vye}
T.~M.~Aliev, K.~Azizi, Y.~Sarac and H.~Sundu,
{\it Determination of the quantum numbers of $\Sigma_b(6097)^{\pm}$ via their strong decays},
Phys. Rev. D \textbf{99} (2019) no.9, 094003.

\bibitem{Wang:2020pri}
Z.~G.~Wang,
{\it Analysis of the $\Omega_b(6316)$, $\Omega_b(6330)$, $\Omega_b(6340)$ and $\Omega_b(6350)$ with QCD sum rules},
Int. J. Mod. Phys. A \textbf{35} (2020) no.07, 2050043.

\bibitem{Agaev:2020fut}
S.~S.~Agaev, K.~Azizi and H.~Sundu,
{\it Newly discovered $\Xi _c^{0}$ resonances and their parameters},
Eur. Phys. J. A \textbf{57} (2021) no.6, 201.

\bibitem{Azizi:2020tgh}
K.~Azizi, Y.~Sarac and H.~Sundu,
{\it $\Lambda_b(6146)^0$ state newly observed by LHCb},
Phys. Rev. D \textbf{101} (2020) no.7, 074026.

\bibitem{Yu:2021zvl}
G.~L.~Yu, Z.~G.~Wang and X.~W.~Wang,
{\it The $1D$, $2D$ $\Xi_{b}$ and $\Lambda_{b}$ baryons},
[arXiv:2109.02217 [hep-ph]].

\bibitem{Burch:2015pka}
T.~Burch,
{\it Heavy hadrons on $N_f=2$ and $2+1$ improved clover-Wilson lattices},
[arXiv:1502.00675 [hep-lat]].

\bibitem{Padmanath:2013bla}
M.~Padmanath, R.~G.~Edwards, N.~Mathur and M.~Peardon,
{\it Excited-state spectroscopy of singly, doubly and triply-charmed baryons from lattice QCD},
[arXiv:1311.4806 [hep-lat]].

\bibitem{Padmanath:2017lng}
M.~Padmanath and N.~Mathur,
{\it Quantum Numbers of Recently Discovered $\Omega^{0}_{c}$ Baryons from Lattice QCD},
Phys. Rev. Lett. \textbf{119} (2017) no.4, 042001.

\bibitem{Crede:2013kia}
V.~Crede and W.~Roberts,
{\it Progress towards understanding baryon resonances},
Rept. Prog. Phys. \textbf{76} (2013), 076301.

\bibitem{Cheng:2015iom}
H.~Y.~Cheng,
{\it Charmed baryons circa 2015},
Front. Phys. (Beijing) \textbf{10} (2015) no.6, 101406.

\bibitem{Cheng:2021qpd}
H.~Y.~Cheng,
{\it Charmed Baryon Physics Circa 2021},
[arXiv:2109.01216 [hep-ph]].

\bibitem{Meng:2022ozq}
L.~Meng, B.~Wang, G.~J.~Wang and S.~L.~Zhu,
{\it Chiral perturbation theory for heavy hadrons and chiral effective field theory for heavy hadronic molecules},
[arXiv:2204.08716 [hep-ph]].

\bibitem{Yang:2021lce}
H.~M.~Yang and H.~X.~Chen,
{\it $P$-wave charmed baryons of the $SU(3)$ flavor $6_F$},
Phys. Rev. D \textbf{104} (2021) no.3, 034037.

\bibitem{Yang:2020zrh}
H.~M.~Yang and H.~X.~Chen,
{\it $P$-wave bottom baryons of the $SU(3)$ flavor $\mathbf{6}_F$},
Phys. Rev. D \textbf{101} (2020) no.11, 114013,
[erratum: Phys. Rev. D \textbf{102} (2020) no.7, 079901].

\bibitem{Yang:2020zjl}
H.~M.~Yang, H.~X.~Chen and Q.~Mao,
{\it Identifying the $\Xi_c^0$ baryons observed by LHCb as $P$-wave $\Xi_c^\prime$ baryons},
Phys. Rev. D \textbf{102} (2020), 114009.

\bibitem{Mao:2020jln}
Q.~Mao, H.~X.~Chen and H.~M.~Yang,
{\it Identifying the $\Lambda_b(6146)^0$ and $\Lambda_b(6152)^0$ as $D$-wave bottom baryons},
Universe \textbf{6} (2020) no.6, 86.

\bibitem{Cui:2019dzj}
E.~L.~Cui, H.~M.~Yang, H.~X.~Chen and A.~Hosaka,
{\it Identifying the $\Xi_{b}(6227)$ and $\Sigma_{b}(6097)$ as $P$-wave bottom baryons of $J^P = 3/2^-$},
Phys. Rev. D \textbf{99} (2019) no.9, 094021.

\bibitem{Chen:2020mpy}
H.~X.~Chen, E.~L.~Cui, A.~Hosaka, Q.~Mao and H.~M.~Yang,
{\it Excited $\varOmega _b$ baryons and fine structure of strong interaction},
Eur. Phys. J. C \textbf{80} (2020) no.3, 256.

\bibitem{Chen:2017sci}
H.~X.~Chen, Q.~Mao, W.~Chen, A.~Hosaka, X.~Liu and S.~L.~Zhu,
{\it Decay properties of $P$-wave charmed baryons from light-cone QCD sum rules},
Phys. Rev. D \textbf{95} (2017) no.9, 094008.

\bibitem{Mao:2015gya}
Q.~Mao, H.~X.~Chen, W.~Chen, A.~Hosaka, X.~Liu and S.~L.~Zhu,
{\it QCD sum rule calculation for $P$-wave bottom baryons},
Phys. Rev. D \textbf{92} (2015) no.11, 114007.

\bibitem{Chen:2015kpa}
H.~X.~Chen, W.~Chen, Q.~Mao, A.~Hosaka, X.~Liu and S.~L.~Zhu,
{\it $P$-wave charmed baryons from QCD sum rules},
Phys. Rev. D \textbf{91} (2015) no.5, 054034.

\bibitem{Yang:2019cvw}
H.~M.~Yang, H.~X.~Chen, E.~L.~Cui, A.~Hosaka and Q.~Mao,
{\it Decay properties of $P$-wave bottom baryons within light-cone sum rules},
Eur. Phys. J. C \textbf{80} (2020) no.2, 80.

\bibitem{Shifman:1978bx}
M.~A.~Shifman, A.~I.~Vainshtein and V.~I.~Zakharov,
{\it QCD and Resonance Physics. Theoretical Foundations},
Nucl. Phys. B \textbf{147} (1979), 385-447.

\bibitem{Reinders:1984sr}
L.~J.~Reinders, H.~Rubinstein and S.~Yazaki,
{\it Hadron Properties from QCD Sum Rules},
Phys. Rept. \textbf{127} (1985), 1.

\bibitem{Balitsky:1989ry}
I.~I.~Balitsky, V.~M.~Braun and A.~V.~Kolesnichenko,
{\it Radiative Decay $\Sigma^+ \to p \gamma$ in Quantum Chromodynamics},
Nucl. Phys. B \textbf{312} (1989), 509-550.

\bibitem{Braun:1988qv}
V.~M.~Braun and I.~E.~Filyanov,
{\it QCD Sum Rules in Exclusive Kinematics and Pion Wave Function},
Z. Phys. C \textbf{44} (1989), 157.

\bibitem{Chernyak:1990ag}
V.~L.~Chernyak and I.~R.~Zhitnitsky,
{\it B meson exclusive decays into baryons},
Nucl. Phys. B \textbf{345} (1990), 137-172.

\bibitem{Brodsky:1997de}
S.~J.~Brodsky, H.~C.~Pauli and S.~S.~Pinsky,
{\it Quantum chromodynamics and other field theories on the light cone},
Phys. Rept. \textbf{301} (1998), 299-486.

\bibitem{Ball:1998je}
P.~Ball,
{\it Theoretical update of pseudoscalar meson distribution amplitudes of higher twist: The Nonsinglet case},
JHEP \textbf{01} (1999), 010.

\bibitem{Ball:2006wn}
P.~Ball, V.~M.~Braun and A.~Lenz,
{\it Higher-twist distribution amplitudes of the $K$ meson in QCD},
JHEP \textbf{05} (2006), 004.

\bibitem{Grinstein:1990mj}
B.~Grinstein,
{\it The Static Quark Effective Theory},
Nucl. Phys. B \textbf{339} (1990), 253-268.

\bibitem{Eichten:1989zv}
E.~Eichten and B.~R.~Hill,
{\it An Effective Field Theory for the Calculation of Matrix Elements Involving Heavy Quarks},
Phys. Lett. B \textbf{234} (1990), 511-516.

\bibitem{Falk:1990yz}
A.~F.~Falk, H.~Georgi, B.~Grinstein and M.~B.~Wise,
{\it Heavy Meson Form-factors From {QCD}},
Nucl. Phys. B \textbf{343} (1990), 1-13.

\bibitem{Arifi:2021orx}
A.~J.~Arifi, D.~Suenaga and A.~Hosaka,
{\it Relativistic corrections to decays of heavy baryons in the quark model},
Phys. Rev. D \textbf{103} (2021) no.9, 094003.

\bibitem{He:2021xrh}
H.~Z.~He, W.~Liang, Q.~F.~L\"u and Y.~B.~Dong,
{\it Strong decays of the low-lying bottom strange baryons},
Sci. China Phys. Mech. Astron. \textbf{64} (2021) no.6, 261012.

\bibitem{Kim:2021ywp}
Y.~Kim, Y.~R.~Liu, M.~Oka and K.~Suzuki,
{\it Heavy baryon spectrum with chiral multiplets of scalar and vector diquarks},
Phys. Rev. D \textbf{104} (2021) no.5, 054012.

\bibitem{Hazra:2021lpa}
A.~Hazra, S.~Rakshit and R.~Dhir,
{\it Radiative M1 transitions of heavy baryons: Effective quark mass scheme},
Phys. Rev. D \textbf{104} (2021) no.5, 053002.

\bibitem{Chen:2022fye}
B.~Chen, S.~Q.~Luo, K.~W.~Wei and X.~Liu,
{\it b-hadron spectroscopy study based on the similarity of double bottom baryon and bottom meson},
Phys. Rev. D \textbf{105} (2022) no.7, 074014.

\bibitem{Polyakov:2022eub}
M.~V.~Polyakov and M.~Praszalowicz,
{\it Landscape of heavy baryons from the perspective of the chiral quark-soliton model},
Phys. Rev. D \textbf{105} (2022), 094004.

\bibitem{Kakadiya:2022zvy}
A.~Kakadiya, Z.~Shah and A.~K.~Rai,
{\it Mass Spectra and Decay Properties of Singly Heavy Bottom-Strange Baryons},
[arXiv:2202.12048 [hep-ph]].

\bibitem{Dong:2022otb}
R.~R.~Dong, N.~Su and H.~X.~Chen,
{\it Fully-strange tetraquark states with the exotic quantum number $J^{PC} = 4^{+-}$},
[arXiv:2206.09517 [hep-ph]].

\bibitem{Yang:1993bp}
K.~C.~Yang, W.~Y.~P.~Hwang, E.~M.~Henley and L.~S.~Kisslinger,
{\it QCD sum rules and neutron proton mass difference},
Phys. Rev. D \textbf{47} (1993), 3001-3012.

\bibitem{Hwang:1994vp}
W.~Y.~P.~Hwang and K.~C.~Yang,
{\it QCD sum rules: $\Delta - N$ and $\Sigma^0 - \Lambda$ mass splittings},
Phys. Rev. D \textbf{49} (1994), 460-465.

\bibitem{Ovchinnikov:1988gk}
A.~A.~Ovchinnikov and A.~A.~Pivovarov,
{\it QCD sum rule calculation of the quark gluon condensate},
Sov. J. Nucl. Phys. \textbf{48} (1988), 721-723.

\bibitem{Jamin:2002ev}
M.~Jamin,
{\it Flavor symmetry breaking of the quark condensate and chiral corrections to the Gell-Mann-Oakes-Renner relation},
Phys. Lett. B \textbf{538} (2002), 71-76.

\bibitem{Ioffe:2002be}
B.~L.~Ioffe and K.~N.~Zyablyuk,
{\it Gluon condensate in charmonium sum rules with three loop corrections},
Eur. Phys. J. C \textbf{27} (2003), 229-241.

\bibitem{Gimenez:2005nt}
V.~Gimenez, V.~Lubicz, F.~Mescia, V.~Porretti and J.~Reyes,
{\it Operator product expansion and quark condensate from lattice QCD in coordinate space},
Eur. Phys. J. C \textbf{41} (2005), 535-544.

\bibitem{Colangelo:1998ga}
P.~Colangelo, F.~De Fazio and N.~Paver,
{\it Universal $\tau_{1/2}$(y) Isgur-Wise function at the next-to-leading order in QCD sum rules},
Phys. Rev. D \textbf{58} (1998), 116005.

\bibitem{Yoshida:2015tia}
T.~Yoshida, E.~Hiyama, A.~Hosaka, M.~Oka and K.~Sadato,
{\it Spectrum of heavy baryons in the quark model},
Phys. Rev. D \textbf{92} (2015) no.11, 114029.

\bibitem{Nagahiro:2016nsx}
H.~Nagahiro, S.~Yasui, A.~Hosaka, M.~Oka and H.~Noumi,
{\it Structure of charmed baryons studied by pionic decays},
Phys. Rev. D \textbf{95} (2017) no.1, 014023.

\bibitem{Ball:2004rg}
P.~Ball and R.~Zwicky,
{\it $B_{d,s} \to  \rho, \omega, K^*, \phi$ decay form-factors from light-cone sum rules revisited},
Phys. Rev. D \textbf{71} (2005), 014029.

\bibitem{Ball:1998kk}
P.~Ball and V.~M.~Braun,
{\it Exclusive semileptonic and rare B meson decays in QCD},
Phys. Rev. D \textbf{58} (1998), 094016.

\bibitem{Ball:1998sk}
P.~Ball, V.~M.~Braun, Y.~Koike and K.~Tanaka,
{\it Higher twist distribution amplitudes of vector mesons in QCD: Formalism and twist - three distributions},
Nucl. Phys. B \textbf{529} (1998), 323-382.

\bibitem{Ball:1998ff}
P.~Ball and V.~M.~Braun,
{\it Higher twist distribution amplitudes of vector mesons in QCD: Twist-$4$ distributions and meson mass corrections},
Nucl. Phys. B \textbf{543} (1999), 201-238.

\bibitem{Ball:2007rt}
P.~Ball and G.~W.~Jones,
{\it Twist-$3$ distribution amplitudes of $K^*$ and $\phi$ mesons},
JHEP \textbf{03} (2007), 069.

\bibitem{Ball:2007zt}
P.~Ball, V.~M.~Braun and A.~Lenz,
{\it Twist-$4$ distribution amplitudes of the $K^*$ and $\phi$ mesons in QCD},
JHEP \textbf{08} (2007), 090.

\bibitem{LHCb:2017uwr}
R.~Aaij \textit{et al.} [LHCb],
{\it Observation of five new narrow $\Omega_c^0$ states decaying to $\Xi_c^+ K^-$},
Phys. Rev. Lett. \textbf{118} (2017) no.18, 182001

\bibitem{Chen:2017gnu}
B.~Chen and X.~Liu,
{\it New $\Omega_c^0$ baryons discovered by LHCb as the members of $1P$ and $2S$ states},
Phys. Rev. D \textbf{96} (2017) no.9, 094015
LaTeX (EU)





















\end{thebibliography}
\end{document}